\documentclass[fleqn,usenatbib]{mnras}

\usepackage{newtxtext,newtxmath}
\usepackage[flushleft]{threeparttable}
\usepackage{lscape}

\usepackage[T1]{fontenc}
\usepackage{multirow}

\DeclareRobustCommand{\VAN}[3]{#2}
\let\VANthebibliography\thebibliography
\def\thebibliography{\DeclareRobustCommand{\VAN}[3]{##3}\VANthebibliography}

\usepackage{graphicx,caption,subcaption}	
\usepackage{amsmath}	
\usepackage{array}
\usepackage[left]{lineno}

\title[The TESS Triple-9 Catalog 2.0]{The TESS Triple-9 Catalog II: a new set of 999 uniformly-vetted exoplanet candidates}

\author[C. Magliano et al.]{
Christian Magliano$^{1,2,3}$\thanks{E-mail: christian.magliano@unina.it},
Veselin Kostov $^{4,5}$,
Luca Cacciapuoti$^{6}$,
Giovanni Covone$^{1,2,3}$,
Laura Inno$^{7}$,
\newauthor
Stefano Fiscale$^{7}$,
Marc Kuchner$^{4}$,
Elisa V. Quintana$^{4}$,
Ryan Salik$^{8,9}$,
Vito Saggese$^{1}$,
John M. Yablonsky$^{9}$,
\newauthor
Aline U. Fornear$^{9}$,
Michiharu Hyogo$^{9,10}$,
Marco Z. Di Fraia$^{9,11}$,
Hugo A. Durantini Luca$^{9,16}$,
\newauthor
Julien S. de Lambilly$^{9}$,
Fabrizio Oliva$^{12}$,
Isabella Pagano$^{13}$,
Riccardo M. Ienco$^{9,1}$,
Lucas T. de Lima$^{9,14}$,
\newauthor
Marc Andrés-Carcasona$^{9,15}$,
Francesco Gallo$^{9,1}$,
Sovan Acharya$^{9,16}$.
\\
$^{1}$Dipartimento di Fisica ``Ettore Pancini'', Università di Napoli Federico II, 80126 Napoli, Italy\\
$^{2}$INFN, Sezione di Napoli, Complesso Universitario di Monte S. Angelo,
Via Cintia Edificio 6, 80126 Napoli, Italy \\
$^{3}$INAF - Osservatorio Astronomico di Capodimonte,
via Moiariello 16, 80131 Napoli, Italy \\
$^{4}$NASA Goddard Space Flight Center, 8800 Greenbelt Road, Greenbelt, MD 20771, USA\\
$^{5}$GSFC Sellers Exoplanet Environments Collaboration, USA\\
$^{6}$European Southern Observatory, Karl-Schwarzschild-Strasse 2 D-85748 Garching bei Munchen, Germany \\
$^{7}$Science and Technology Department, Parthenope University of Naples, Centro Direzionale, Isola C4, I-80143 Naples, Italy \\
$^{8}$Staples High School, 70 North Ave, Westport, CT 06880, USA\\
$^{9}$Citizen Scientist, Planet Patrol Collaboration\\
$^{10}$ School of Science and Engineering, Meisei University, 2-1-1 Hodokubo, Hino, Tokyo 191-0042, Japan\\
$^{11}$Long Tail Research, Percy Street, Oxford, OX4 3AD, UK\\
$^{12}$Istituto di Astrofisica e Planetologia Spaziali (IAPS/INAF), 00133 Rome, Italy\\
$^{13}$INAF - Osservatorio Astrofisico di Catania, 95123 Catania, Italy\\
$^{14}$Geoscience Department, University of Aveiro. Campus Santiago, 3810-193, Aveiro\\
$^{15}$Institut de Física d’Altes Energies (IFAE), Barcelona Institute of Science and Technology, E-08193 Barcelona, Spain\\
$^{16}$SA Citizen Science Group-Ignited Minds VIPNET Club, India
}
\date{Accepted XXX. Received YYY; in original form ZZZ}

\pubyear{2015}

\begin{document}
\label{firstpage}
\pagerange{\pageref{firstpage}--\pageref{lastpage}}
\maketitle

\newcommand{\ntics}{999}
\begin{abstract}
The Transiting Exoplanet Survey Satellite (TESS) mission is providing the scientific community with millions of light curves of stars spread across the whole sky. 
Since 2018 the telescope has detected thousands of planet candidates that need to be meticulously scrutinized before being considered amenable targets for follow-up programs. We present the second catalog of the Plant Patrol citizen science project containing \ntics{} uniformly-vetted exoplanet candidates within the TESS ExoFOP archive. The catalog was produced by fully exploiting the power of the Citizen Science Planet Patrol project. We vetted TESS Objects of Interest (TOIs) based on the results of Discovery And Vetting of Exoplanets (\texttt{DAVE}) pipeline. 
We also implemented the Automatic Disposition Generator, a custom procedure aimed at generating the final classification for each TOI that was vetted by at least three vetters. The majority of the candidates in our catalog, 752 TOIs, passed the vetting process and were labelled as planet candidates. We ruled out $142$ candidates as false positives and flagged $105$ as potential false positives. Our final dispositions and comments for all the planet candidates are provided as a publicly available supplementary table.
\end{abstract}




\section{Introduction}
Over the years, more than 5,000 exoplanets have been discovered as a result of a series of ground- and space-based exoplanet-hunting missions, with thousands more still awaiting confirmation\footnotemark{}. The development of extensive catalogs of confirmed exoplanets 
is a fundamental step to shed light on the planetary formation processes and provide clues as to whether the Solar System is unique in the Galaxy \citep[e.g.,][]{2021MNRAS.500.1313B}.
Among the different exoplanet discovery techniques, the transit method has proven to be the most fruitful, leading to the discovery of $\sim77\%$ of all currently known exoplanets\footnotemark[\value{footnote}].\footnotetext{{https://exoplanetarchive.ipac.caltech.edu/}} In fact, the Kepler mission \citep{2010Sci...327..977B} alone has found over 2,700 transit-like signals that were later verified as true planets along with many candidates still waiting further examination \citep{2022LPICo2687.3027L}. Since its launch in 2018, the Transiting Exoplanet Survey Satellite (TESS; \citealt{2015JATIS...1a4003R}) space-based mission has observed $\sim85\%$ of the entire sky, collecting the light curves of $\sim$ 200,000 pre-selected stars at $2$-minute cadence. Furthermore, TESS also acquired a series of Full Frame Images (FFIs) at 10 and 30 minute cadences, with the goal of expanding the transit search to the entire sky; since September 2022 (the mission's second extension), FFIs have being acquired at a cadence of $200$ s.
At the time of writing, TESS has detected almost 6,000 TESS Object of Interest (TOIs) while $\sim $ 10,000 are expected to be found in the FFIs within the primary mission duration \citep{2018ApJS..239....2B}. According to the ExoFOP-TESS archive\footnote{https://exofop.ipac.caltech.edu/tess/}, 
at December 2022, 277 candidates out of the currently-known 5,887 TOIs have been validated to date by follow-up measurements. 

Detecting a transit-like signal in the light curve of a distant star is not sufficient to confirm the discovery of an exoplanet. Several astrophysical sources (e.g., eclipsing binary stars, stellar spots and/or pulsations; \citealt{2018arXiv181008689C}) or instrumental artefacts (e.g., jitter noise and momentum dumps) can mimic a transit-like signal in the light curve of the observed target leading to a \textit{false positive} detection. In light of the many potential false positive scenarios that affect the photometric transit method, a planet candidate has to be carefully examined before being promoted as a suitable target for spectroscopic follow-up observations aimed at its confirmation as a bona-fide planet. Precision radial velocity (PRV; \citealt{1996A&AS..119..373B,2004A&A...423..385P}) measurements are challenging, time-consuming, and achievable by only a handful of instruments. 

The vetting procedure is one of the key steps in the process of confirming the planetary origin of a transit feature found in the light curve of a star. A catalog of uniformly-vetted transiting planet candidates is essential to optimize spectroscopic follow-up observations by promoting targets for which common false positive scenarios have been already ruled out. Moreover, the vetting procedure enables statistical validation of planet candidates for which no PRV measurements are feasible. Finally, complementary human vetting also provides the opportunity to create a knowledge base for machine learning approaches aiming to automate the entire vetting process.

Several automated vetting pipelines have been developed over the years to tackle the issue of false positives in transit photometry. 
For example, the \texttt{AUTOVETTER} \citep{2015ApJ...806....6M}, \texttt{ROBOVETTER} \citep{2016ApJS..224...12C} and \texttt{SIDRA} \citep{2016MNRAS.455..626M} pipelines are decision-tree based machine learning codes trained on massive human-inspected data sets to produce uniformly-vetted catalogs of planet candidates discovered from the Kepler mission. 

Deep learning algorithms have been trained to identify planet candidates in both Kepler and TESS light curves. These work both as likelihood-based rankers (\citealt{2018AJ....155...94S}) or binary classifiers (e.g. \citealt{Olm2021}). Since the innovative and high-performance approach provided by these models, vetting efforts have shifted towards deep learning (DL) methods. Despite the fact that DL models usually outperform traditional machine learning methods, they come with certain drawbacks. Most notably, DL models are computationally expensive and the results they produce are sometimes difficult to interpret \citep{2017arXiv170808296S}.

Apart from models based on neural networks, pipelines such as \texttt{VESPA} \citep{2012ApJ...761....6M} and \texttt{TRICERATOPS} \citep{2021AJ....161...24G} evaluate the Bayesian probability that a signal is a false positive based on the shape of the light curve as well as the stellar parameters of the nearby sources within the aperture mask used to extract the light curve. Furthermore, once a certain false positive threshold value is set, these algorithms allow to statistically validate a signal as a true planet. The Discovery And Vetting of Exoplanets (DAVE) \cite{2019AJ....157..124K} vetting pipeline determines whether a transit-like signal is caused by a planet candidate or is a false positive by testing the candidate at both the pixel and light curve levels. Building upon methods used for vetting exoplanet candidates from the Kepler mission, \texttt{DAVE} was designed to analyze transit photometry from the K2 mission, and later modified to work with TESS light curves as well \citep{2019AJ....158...32K,2020AJ....160..116G}.

It is important to note that none of these pipelines can completely replace visual human inspection. Automatic pipelines, for example, can fail to correctly classify signals with low signal-to-noise ratio (SNR) (e.g., small planets with long periods), that are dominated by stellar variability, or that are plagued by various systematic effects and instrumental artefacts. Furthermore, different planet search and/or vetting pipelines use different methods to extract and process the raw data. For example, \cite{2019AJ....157..124K} demonstrated that nearly one in every three K2 planet candidates has insufficient SNR across all available light curve sets to provide a reliable classification (i.e., planet candidate or false positive). Hence, all automated vetting pipelines come with inherent data-processing and data-analyzing biases and peculiarities, making complementary human inspection not only recommended but also essential.

Traditionally, complementary human vetting is typically done by a small group of professional astronomers. However, the ever-increasing number of exoplanet candidates in need of careful examination makes this approach impractical. Vetting hundreds of targets by a handful of scientists may take months; unforeseen biases may emerge unless a clear workflow is defined within the team at the start of the work, and strictly adhered to (e.g. \citealt{2018ApJS..235...38T}), and an intuitive, interactive, user-friendly vetting platform is used by all vetters.

Citizen science is a powerful way of doing science that is becoming increasingly popular due to new collaboration tools. It offers the opportunity to address the human vetting bottleneck by harnessing the expertise and enthusiasm of amateur astronomers. For example projects like Planet Patrol \citep{2022PASP..134d4401K}, Planet Hunter TESS (PHT) \citep{2020MNRAS.494..750E}, Exoplanet Explorers \citep{Christiansen_2018} and Disk Detective \citep{Kuchner_2016}, hosted by the Zooniverse platform \citep{2008MNRAS.389.1179L}, helped scientists achieve, in a few weeks, results that would have otherwise taken years to complete. 

Planet Patrol is a citizen science project designed to assist with the vetting workflow of TESS planet candidates based on the automated results and dispositions produced by the \texttt{DAVE} pipeline \cite{2022PASP..134d4401K}. 
After the first stage of the project was completed on Zooniverse, several citizen scientists expressed interest in continuing assisting the scientific core team with the vetting efforts. Under the guidance of members from our core science team, these \textit{``superuser''} volunteers were trained to classify TESS planet candidates by critically interpreting and analyzing the entire output from \texttt{DAVE}. The superusers became an integral part of the team and played an essential role in our first TESS Triple 9 Catalog (\citealt{2022MNRAS.513..102C}, Paper I hereafter), where they assisted with the vetting of $999$ TOIs, classifying $709$ of them as planet candidates.

In this work we present the continuation of our vetting efforts, in the form of a catalog of \ntics{} uniformly-vetted TESS planet candidates detected by the Science Processing Operations Center pipeline (SPOC, \citealt{2016SPIE.9913E..3EJ}) and Quick Look Pipeline (QLP, \citealt{2020RNAAS...4..204H}) pipelines. We utilize the same workflow as used in Paper I and introduce several new vetting tools and diagnostics.

The outline of the paper is the following: in Section \ref{sec:method} we discuss the workflow adopted to uniformly vet \ntics{} candidates within the TESS database, including the new implementations with respect to Paper I. In Section \ref{sec:planetpatrol} we highlight the details of the Planet Patrol Project and how it helped in carrying out this work. The catalog and its details are discussed in Section \ref{sec:catalog}. Finally we summarize our conclusions in Section \ref{sec:conclusions}.

\section{Method}
\label{sec:method}
We have conducted a uniform vetting of \ntics{} TOIs by means of the \texttt{DAVE} pipeline. \texttt{DAVE} utilizes a two-step vetting process for each TESS sector where the target has been observed, namely a pixel-level photocenter analysis and a flux-based analysis at the light curve level. 
The pipeline vets both the SPOC short-cadence and the FFI long-cadence TESS data, using the ``Corrected Flux'' \texttt{eleanor} light curves \citep{2019PASP..131i4502F} for the latter as-is, i.e. without further detrending or post-processing.
\texttt{DAVE} uses the target's TIC ID, transit ephemeris, depth and duration as provided by the publicly-available ExoFOP website. For completeness, we outline the main products of \texttt{DAVE} below; for further details we refer the reader to \cite{2019AJ....157..124K}.

\begin{figure}
    \centering
    \includegraphics[width=0.5\textwidth]{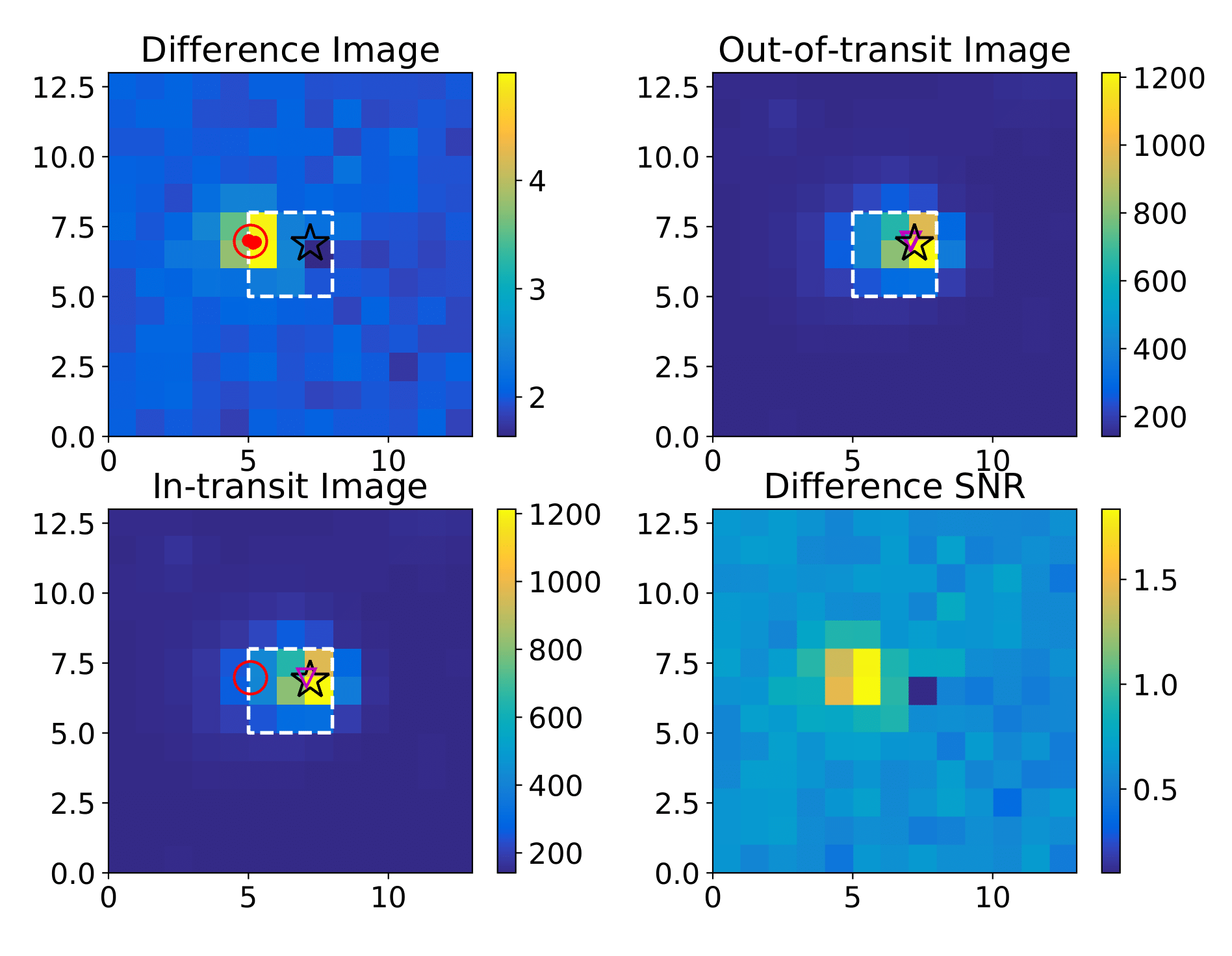}
    \caption{\texttt{DAVE} photocentre analysis of planet candidate TIC 253917293.01. The dashed white contour is the aperture mask used to extract the light curve, the star symbol represents the catalog position of the target, the purple triangle is the measured average out-of-transit photocentre, the small red dots represent the position of the individual photocentres and the large red circle represents the measured overall difference image photocentre. Upper left: the difference image; upper right: the average out-of-transit image; lower left: the average in-transit image; lower right: signal-to-noise ratio of the mean difference image. The color bar indicates the number of electrons/sec for each of the aforementioned cases. The difference image clearly shows a centroid offset and no artefacts. Hence, we rule out this TIC as a false positive due to a clear centroid offset.}
    \label{fig:CO_example}
\end{figure}

\begin{figure}
    \centering
    \includegraphics[width=0.45\textwidth]{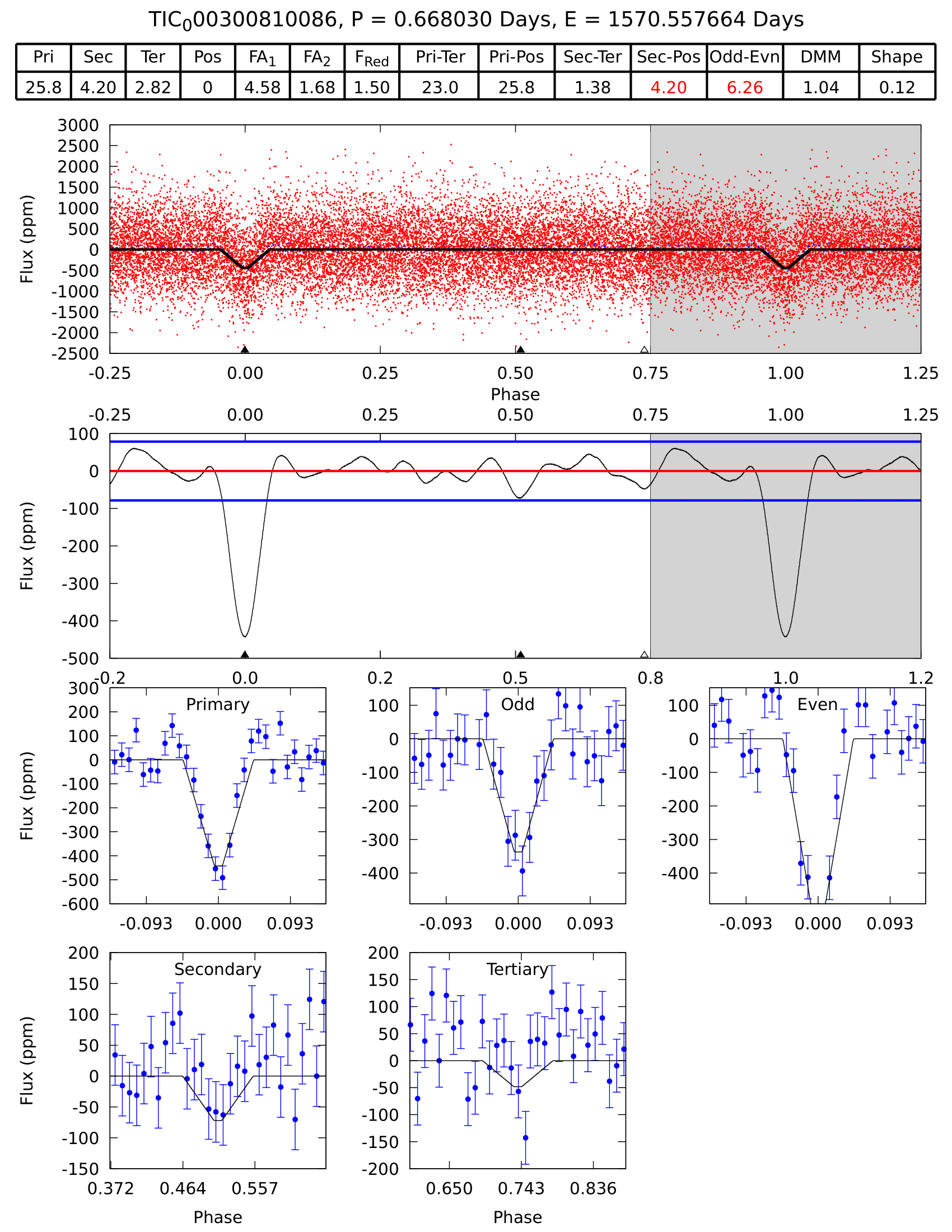} 
    \caption{\texttt{DAVE} Modelshift analysis of planet candidate TIC 300810086.01. The first panel of the Modelshift shows the phase-folded light curve along with the best-fit trapezoid transit model (black line); the second panel depicts the light curve convolved with the transit model and the scatter level (horizontal blue lines); the lower panels shows zoom-ins on the primary and secondary events, the odd and even primary events, along with any tertiary or positive events. The uppermost table displays the statistical significance of the aforementioned features, red-colored if the pipeline flags an issue as significant. The Modelshift shows a prominent V-shaped primary and a more than $6\sigma$ significant odd-even difference. Hence, we rule out this TIC as a false positive.}
    \label{fig:VShape+OED_example}
\end{figure}

\begin{figure*}
     \centering
     \begin{subfigure}[b]{\columnwidth}
         \centering
         \includegraphics[width=\columnwidth]{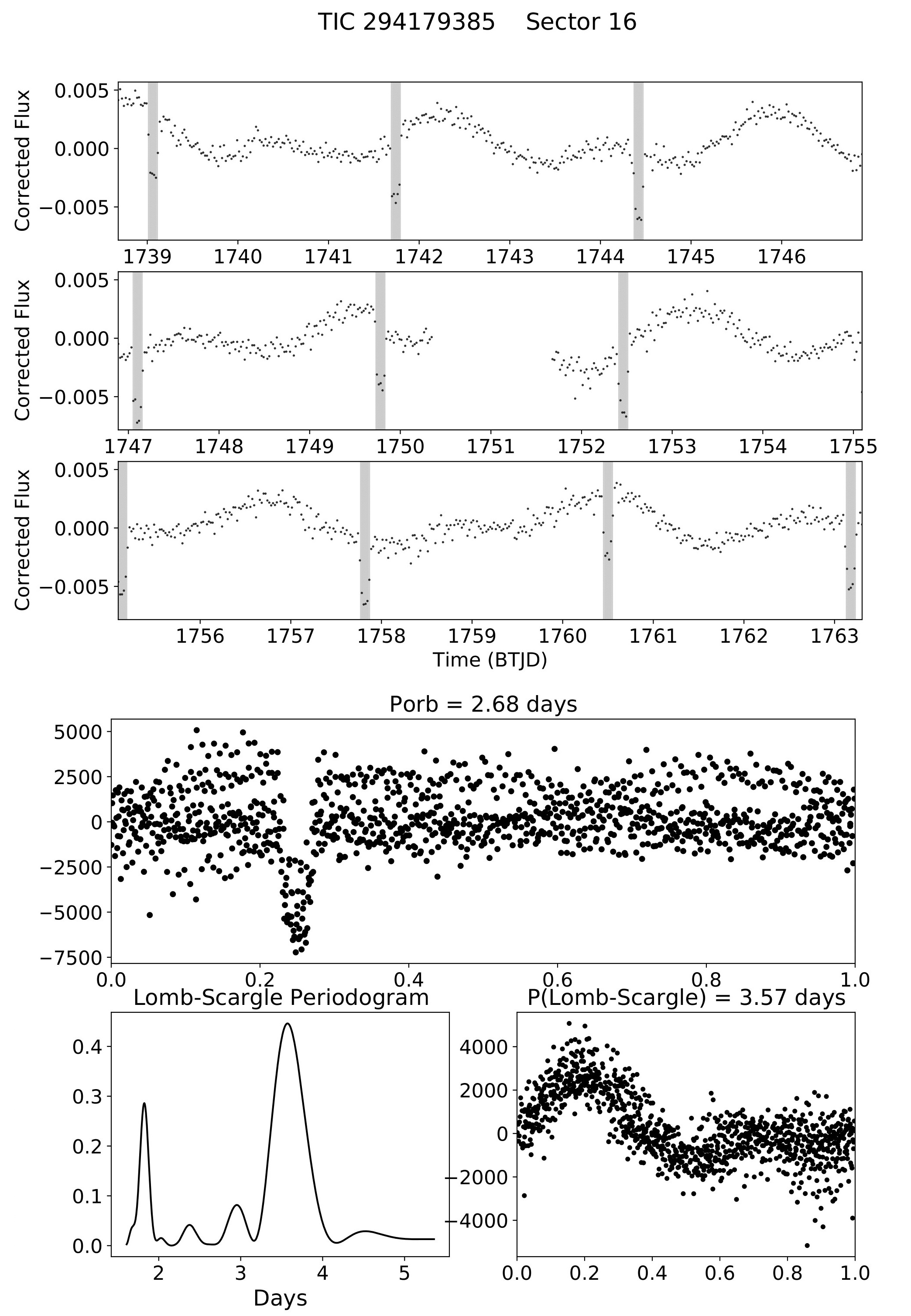}
        \label{fig:lc 294179385}
     \end{subfigure}
     \hfill
     \begin{subfigure}[b]{\columnwidth}
         \centering
         \includegraphics[width=\columnwidth]{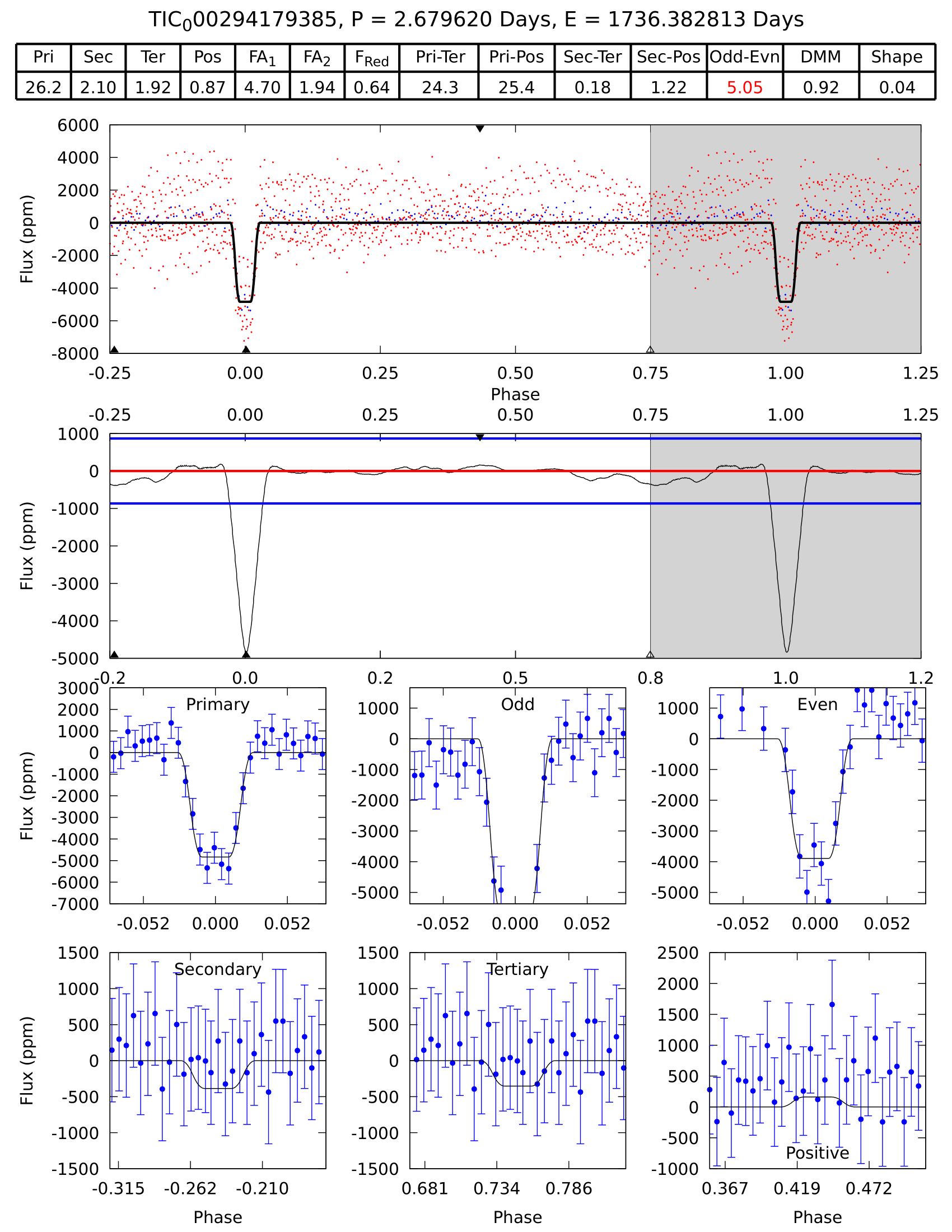}
         \label{fig:mod 294179385}
     \end{subfigure}
        \caption{TIC 294179385, observed in sectors 14, 15, 16 and 41 at 30 minute cadence, is a planet candidate detected by the QLP pipeline with a period $P=2.68$ days. The eleanor light curve (upper left panel) shows a prominent stellar variability with a period of $3.57$ days according to our LS analysis (lower left panel). Unlike the example shown in Fig.\ref{fig:TIC 355637190} the $P_{LS}$ is not suspicious of a BEER scenario because the modulation period is different from the orbital period, and the variability is likely caused by starspots. The aperture mask used for the light curve extraction includes a number of field stars that are bright enough to produce the modulation signal; one of these stars, TIC 294179389, is brighter than the target itself. Hence, the observed light curve modulation can be produced by a nearby field star. Importantly, the prominent stellar variability tricks the Modelshift (right panel) module into flagging the target as a false positive due to a nominal OED. The human vetters inspect the light curve, note the position of the transits with respect to the light curve modulations, and compare the out-of transit baseline level of panels ``Odd'' and ``Even''. After a comprehensive group discussion, we overrule the Modelshift OED disposition and mark the target as a genuine planet candidate.}
        \label{fig:TIC 294179385}
\end{figure*}

\begin{figure*}
     \centering
     \begin{subfigure}[b]{\columnwidth}
         \centering
         \includegraphics[width=\columnwidth]{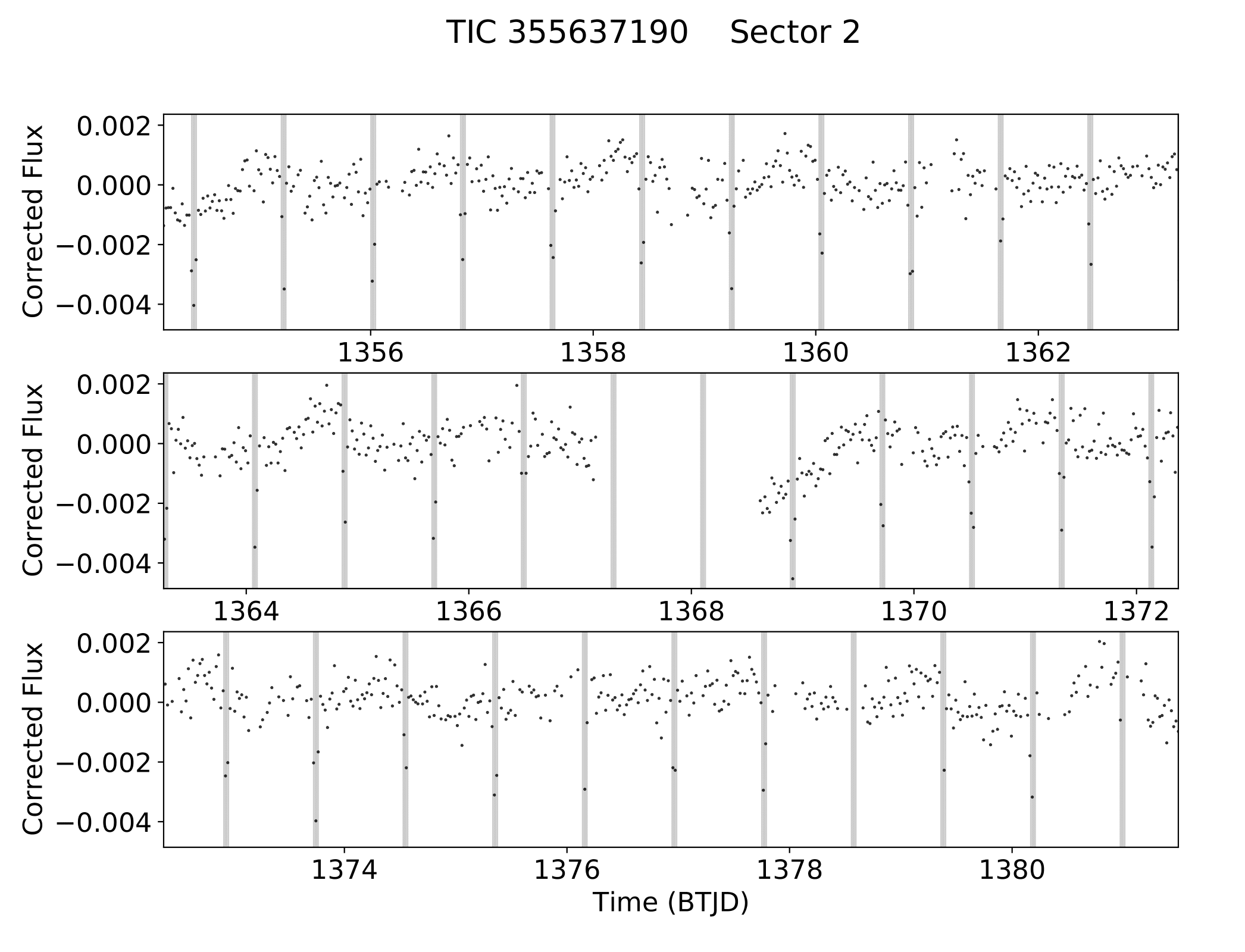}
         \label{fig:lc 355637190_0}
     \end{subfigure}
     \hfill
     \begin{subfigure}[b]{\columnwidth}
         \centering
         \includegraphics[width=\columnwidth]{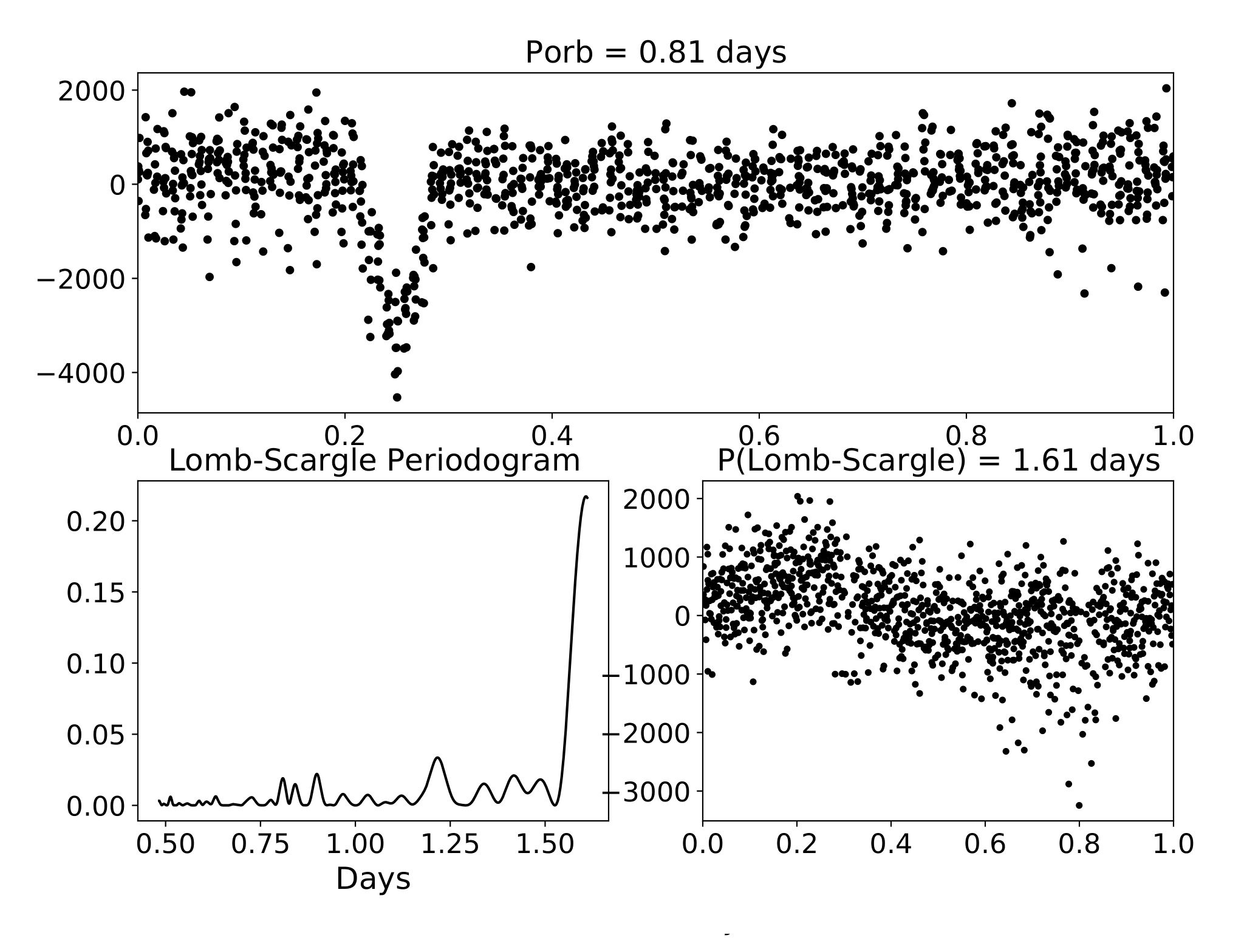}
         \label{fig:mod 355637190_1}
     \end{subfigure}
        \caption{TIC 355637190, observed in sectors 1, 2, 28 and 29 at 30 minute cadence, is a planet candidate detected by the QLP pipeline with a period of $P=0.81$ days. Its light curve (left panel) shows a clear variability, emphasized by the LS periodogram (right panel), with period $P_{\rm LS}=1.61$ days, twice the detected planetary period. This is suggestive of synchronous ellipsoidal variations over two orbital periods in a close binary system. Together with the potential V-shaped transit, it implies that TIC 355637190 does not completely pass our vetting workflow. Hence, even though there are no clear red flags from the \textit{centroids} and \textit{Modelshift} modules, we classify this target as a potential false positive due to the potential ellipsoidal variations.}
        \label{fig:TIC 355637190}
\end{figure*}

\begin{enumerate}
    \item The \textit{centroids} module generates a difference image by subtracting the overall in-transit image from the corresponding out-of-transit image for each transit and for each sector. Then, for each transit, the code calculates the photocenter of the light distribution by fitting to the difference image the TESS Pixel Response Function (PRF) and a Gaussian point-spread function (PSF). Finally, the overall position of the photocenter for a particular sector is computed by taking the average over all the transit events detected in that sector. We note that the centroid difference images created by \texttt{DAVE} can be difficult to interpret when the SNR is low or there are significant artefacts. In such cases the centroid measurements can be unreliable (flagged as ``UC'', for ``Unreliable Centroid'' in our catalog) and the corresponding automated photocenter disposition might be incorrect. For example, if some of the individual difference images exhibit prominent systematics, the calculated average photocenter position may be affected to the point of \texttt{DAVE} flagging the candidate as a false positive due to a nominal centroid offset. Thus it is important for a human vetter to inspect the individual difference images, the corresponding photocenter measurements, and the average difference image, and evaluate the reliability of the automated photocenter dispositions provided by \texttt{DAVE}. The vetter is trained to distinguish between valid difference images and photocenter measurements, and those that could be affected by instrumental and/or computational systematics. The vetter ignores the poor measurements and makes a final decision based on the reliable photocenters. For example, if there is a clear Centroid Offset (flagged as  ``CO'') with respect to the catalog position of the target star, and there are no obvious systematics that might affect the measurements, then the candidate is flagged as a False Positive (FP) (see Fig. \ref{fig:CO_example}). 
    
    \item The \textit{Modelshift} module uses the phase-folded light curve along with the best-fit trapezoid transit model to evaluate the significance of the primary signal together with any secondary and tertiary signals, as well as of potential Odd-Even Difference (OED) between consecutive transits. This module determines whether the source of the signal is consistent with an eclipsing binary system instead of a transiting planet. For example, if there is a significant secondary eclipse at any phase other than zero or an OED, \texttt{DAVE} flags the target as a false positive (see Fig. \ref{fig:VShape+OED_example}). We note that since \texttt{DAVE} uses the ``Corrected Flux'' \texttt{eleanor} light curves without further processing, highly variable stars that were observed in long-cadence only can trick the pipeline by mimicking an OED. Figure \ref{fig:TIC 294179385} shows an example of this for the case of TIC 294179385, which is considered a false positive by \texttt{DAVE} because of the nominal OED but was labelled as a Planetary Candidate (PC) after human inspection. Thus the human vetter has to inspect the output of the Modelshift module and decides whether the detected features are genuine, also paying close attention to (i) the shape of the signal (whether it is U- or V-shaped); (ii) the depth of the primary signal (with respect to the stellar radius as provided by ExoFOP); (iii) the depth of the secondary signal (with respect to a typical expected depth of planet occultation, on the order of a few hundred parts-per-million); and (iv) the overall shape and amplitude of the light curve variability both in- and out-of-transit.
    
    \item The variability is evaluated by the human vetter together with \texttt{DAVE's} Lomb-Scargle (LS) analysis \citep{1976Ap&SS..39..447L,1982ApJ...263..835S} of the transit-masked light curve. This submodule provides a quantitative and qualitative criteria to evaluate the presence of possible light curve modulations (LCMOD) due to intrinsic and/or rotational variability. If the detected modulations have the same (or half or double) period of the detected transit-like signal, they could be the result of gravitational (beaming effect and tidal ellipsoidal distortion) and/or atmospheric (reflected light and thermal emission) effects in a close binary star system \citep{1993ApJ...419..344M,2011MNRAS.415.3921F,2017PASP..129g2001S}. This particular scenario is usually referred to  as a BEaming, Ellipsoidal, Reflection Binary (BEER) binary. Whenever we found any suspicious modulations strictly related to the orbital detected period we flagged the target using comments such as ellipsoidal variations ('EV') or synchronous variations (``synch''). An example of a synchronous scenario is shown in Fig. \ref{fig:TIC 355637190}.
\end{enumerate}

Overall, while \texttt{DAVE} produces automated dispositions for each target, we mandate complementary human supervision for all targets due to the likelihood of systematics that can affect the pipeline's classification. Importantly, our human vetters can override \texttt{DAVE}'s disposition and ultimately have the final word -- any target in our catalog that exhibits potential signs of concern has been subjected to rigorous group discussions. 

Aside from the vetting dispositions, for each TOI we keep track of any noteworthy features using pre-defined acronyms and free-text comments as described in Table \ref{tab:abbreviations}. As described below, we also updated the workflow presented in Paper I by introducing new diagnostic tests that are useful for the most challenging cases.
\begin{table*}
\begin{threeparttable}
\centering

\begin{tabular}{p{0.10\linewidth}  p{0.20\linewidth}  p{0.60\linewidth}}

\hline

Abbreviation\rule{0pt}{3ex}    & Meaning & Description \\     \hline \hline

Disposition\rule{0pt}{3ex} & &      \\ \hline

PC & Planetary Candidate & A TOI that passed all vetting tests. \\
pFP & potential False Positive & A TOI that does not completely pass the vetting tests.\\
FP & False Positive & A TOI that does not pass the vetting tests. \\ \hline \hline

Comments \tnote{†}\;\rule{0pt}{3ex} &  &       \\ \hline
CO & Centroid Offset & The \textit{centroids} module shows a statistical offset of the photocenter. It indicates that the target star is not the source of the investigated signal. \\ \\
UC & Unreliable Centroids & The \textit{centroids} module is not reliable due to the difference images is too noisy. It is mainly caused by stray light, bright field stars or very weak signals. \\ \\
OED & Odd-Even Difference & The \textit{Modelshift} shows a statistically significant difference between odd and even eclipses. It usually indicates an eclipsing binary star.\\ \\
Vshape & V-shaped & The \textit{Modelshift} highlights that the shape of the transit is V-like and not U-shaped as expected from a typical planetary transit. It might indicate an eclipsing binary star. Indeed, the transit of a planet produces a sharp ingress, a flat bottom, and a sharp egress. An eclipsing star mostly produces gradual ingress and egress due to the two objects have comparable sizes. \\ \\
LCMOD & Light Curve MODulation & Both \textit{Modelshift} and \textit{Lomb-Scargle} periodogram indicate oscillations in the starlight due to intrinsic and/or rotational variability that are not synchronized with the orbital period. These can be produced by either the target itself of by a nearby field star that falls in the aperture used to extract the light curve. Such lightcurves are generally not indicative of a potential false positive. \\ \\
BEER & BEaming Ellipsoidal Reflection binary system & A close binary star system whose gravitational and atmospheric interactions cause periodic modulations of the light curve.\\ \\
EV/synch & Ellipsoidal Variations/ synchronous  & The \textit{Lomb-Scargle} periodogram highlights LCMOD with the same (or half or double) period of the detected transit. This might indicate a BEER scenario, thus a false positive.\\ \\
FSCP & Field Star in Central Pixel & There is at least one unresolved source within the same pixel of the target (i.e. $<21''$) that is bright enough to contaminate the detected signal. In the worst case, this source might be the true source of the signal. \\ \\
FSOP & Field Star in Other Pixel & There is at least one resolved source within the aperture mask used to extract the observed light curve that is bright enough to contaminate the detected signal. In the worst case, this source might be the true source of the signal. If it is the case we will rule out the target as a FP due to a CO. \\ \\
TD & Too Deep & The transit is particularly deep ($\gtrsim 2.5-3\%$) that might be the result of an eclipsing binary system.\\ \\
NT & No Transit & The \texttt{eleanor} light curve does not show any transit-like signals for QLP-detected TOIs.\\ \\
SS & Significant Secondary & The \textit{Modelshift} shows a statistically significant secondary. A secondary eclipse is typical of an eclipsing binary star. If this is the case the SS is located at half phase. \\ \\
LOWSNR & Low Signal to Noise Ratio & The signal-to-noise ratio of the expected transits is too low for a reliable inspection.\\ \\
HPMS & High Proper Motion Star & The star exhibits a high proper motion as after consulting the SIMBAD archive.\\ \\
AT & Additional Transits & The \textit{Modelshift} shows additional transits in the phase curve. They could be caused by other planets within the system not yet detected.\\ \\
\hline
\end{tabular}
\begin{tablenotes}
\item[†] Each of the comments can be preceded by a 'p' which stands for 'potential'. It is used when the vetter is not fully convinced of that specific flag. 
\end{tablenotes}
\end{threeparttable}
\caption{Table of the acronyms used in this work. This is a resizing of the Table 1 of the Paper I. The reader can find the most updated and complete list of all abbreviations used in our worfklow at https://exogram.vercel.app/dictionary.}
\label{tab:abbreviations}
\end{table*}

\subsection{Ancillary information}
In many cases a target's light curve or target pixel files are affected by prominent systematics and/or the detected transits have a low SNR compared to the baseline variability. This complicates the vetting procedure and can even make it unreliable altogether. To address this issue and confirm or dispel any concern, we use additional information beyond that provided by \texttt{DAVE}. For instance, the vetter manually checks whether the aperture mask used to extract the light curve includes nearby field stars that are bright enough to contaminate the inspected signal. Below we briefly discuss new diagnostics that have been used in this work and we are currently implementing in \texttt{DAVE}, to provide the vetters with a self-consistent tool without asking them to manually seek this ancillary information. 

\subsubsection{Unresolved sources}
TESS has a large pixel scale, 
about $21^{\prime\prime}$/pixel, with a focus-limited PSF. Hence, the flux measured in a single pixel might be contaminated by nearby background or foreground field sources. Based on our experience with \texttt{DAVE} and TESS data, and depending on the particular target and sector, measuring a reliable photocenter offset of ${\sim5-10}$ arcsec (${\sim 0.25-0.5}$ pixels) is relatively straightforward. In contrast, a bona-fide offset of ${\sim1-2}$ arcsec (${\sim 0.05-0.1}$ pixels) is extremely challenging to measure. Thus even if the photocenter module of \texttt{DAVE} does not measure a significant CO, there might still be sufficiently bright field sources that contaminate the target's light curve and/or are too close to the target to reliably rule out as potential source of the detected transit-like signals. 

In the former case, the additional light dilutes the transits, resulting in an underestimated planet radius \citep{2015ApJ...805...16C}. 
To account for this effect, we consult stellar catalogs (e.g., SIMBAD; \citealt{2000A&AS..143....9W}, GAIA EDR3; \citealt{2021A&A...649A...1G}) to check whether known sources fall within the immediate vicinity of the target. Based on the transit depth ($\delta$) and magnitude difference between the target and resolved nearby field stars, the vetter would then investigate whether these alleged sources could have produced the observed transit signal. For a given target of magnitude $m_0$, we considered a threshold $\Delta mag \, =\,  -2.5\log_{10}\delta$. Thus only sources with a magnitude $m_*$ such that $|m_*-m_0|<\Delta mag$ could produce a signal with the same depth of the one observed. The scientific core team provides the vetters with the $\Delta mag$ for each target. If some unresolved stars falls within the same pixel of the target, the vetter adds a comment `FSCP' (Field Stars in Central Pixel). For completeness, the vetter will also flag a 'FSOP' (Field Stars in Other Pixel) whether a bright enough source falls within the aperture mask. 
This is done for the sake of completeness, but it is not a sufficient reason to rule out the target as a false positive. This check is time-consuming and does not need the critical faculties provided by human inspection.
In the future we plan to provide the vetters with a simple tool that, by performing a GAIA DR3 query, returns all the stars within $5$ pixels from the target. It will also mark those sources within the same TESS pixel ($<21^{\prime\prime}$) and colour each one according to their GAIA DR2 magnitude. The pipeline will then automatically flag any source inside and outside the target's pixel that is bright enough to cause the observed dips in the light curve.

\subsubsection{The background flux}
\begin{figure*}
    \centering
    \includegraphics[width=\textwidth]{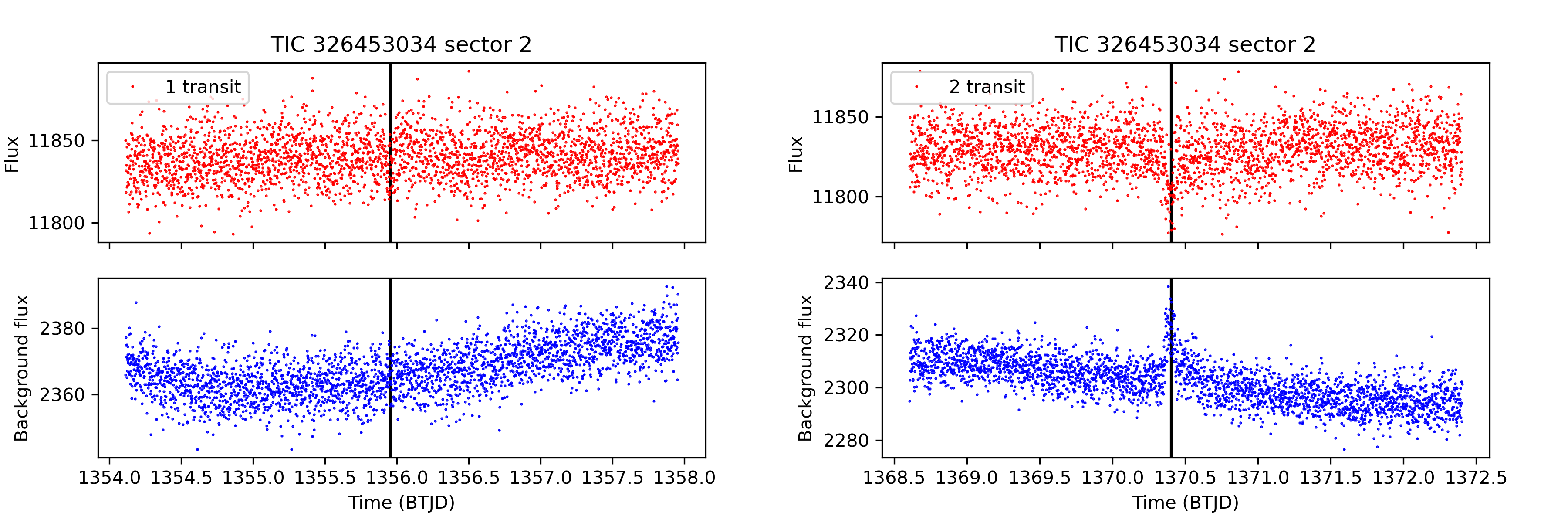}
    \caption{The background flux analysis of planet candidate TIC 326453034 observed in sector 2. With an orbital period of about 14 days, the candidate produced only two potential transits. The upper panels show a zoom-in of the simple aperture photometry flux (red) around the time of the transit. The lower panels show the simple aperture photometry background flux (blue) within the aperture mask in the same time window. The black vertical line indicates the time of the transit for the first (left) and the second (right) transit. The first detected transit has low SNR and the background flux does not exhibit obvious discontinuities. In contrast, the second transit is much better defined, but the background flux shows a sudden spike at the time of the transit. Hence we conclude that this exoplanet candidate is a potential false positive caused by background systematics.}
    \label{fig:bkg_fp}
\end{figure*}

TESS is in a stable, highly elliptical high-Earth orbit in a 2:1 resonance with the Moon. This orbital path ensures maximum sky coverage while minimizing the number of obstructions during data acquisition \citep{2013arXiv1306.5333G}. However this orbital path produces strong contamination in the TESS FFIs, 
mainly from zodiacal light and scattered light from solar system objects \citep{2015ApJ...809...77S}. Hence, the background flux of TESS FFIs varies over the course of the $\sim 27$-day observational window. 
To account for this, we inspect a 4-day long section of the background flux centered on the time of the transit. This helps the vetter determine whether the transit signal seen in the light curve coincides with any background events. In fact, if there is a sudden change in the background flux at or near the time of the transit, it may introduce spurious signals into the light curve mimicking or distorting the transit. 
Thus for each detected transit, we check both the light curve and the background flux in the vicinity of the transit time. If unusual features and/or discontinuities appear in the background during a particular transit, the vetter will flag it as a potential issue. A clear example of a false positive signal due to systematics in the flux background is shown in Fig. \ref{fig:bkg_fp}.

\subsubsection{Pixel Level light curve}

Inspired by the \texttt{LATTE}\footnote{https://github.com/noraeisner/LATTE} pipeline developed within the Planet Hunters TESS project, we decided to include in our workflow a Pixel Level light curve (PLL) analysis. The PLL plot shows the light curve for each individual pixel of the corresponding target pixel file. For further information we refer the reader to \cite{2022ascl.soft05006E}.  We inspect the light curve for each pixel in the field of view, and try to determine whether the transit occurs in the vicinity of the target or originates from another pixel that hosts another star -- yet missed by DAVE's photocenter analysis. This additional layer of scrutiny has proven to be very useful in cases where DAVE's photocenter measurements were unreliable or difficult to interpret. For example, in some cases the scatter in the individual photocenter measurements can be so large that it is practically impossible to distinguish between reliable and spurious measurements. This usually occurs when the individual difference images exhibit a complex pattern (or simply look like random noise) instead of a single bright spot superimposed on a uniform dark background. This is often due to low SNR transits caused by either (i) the presence of nearby field stars that are much brighter than the target itself (and/or are highly variable); or (ii) when the true source of the signal is next to a much brighter target star. In these cases, the PLL analysis helps determine whether some of the detected transits are affected by systematic effects and/or artefacts, and ideally pinpoint the source of the signal. 

\begin{figure*}
    \centering    \includegraphics[width=\textwidth]{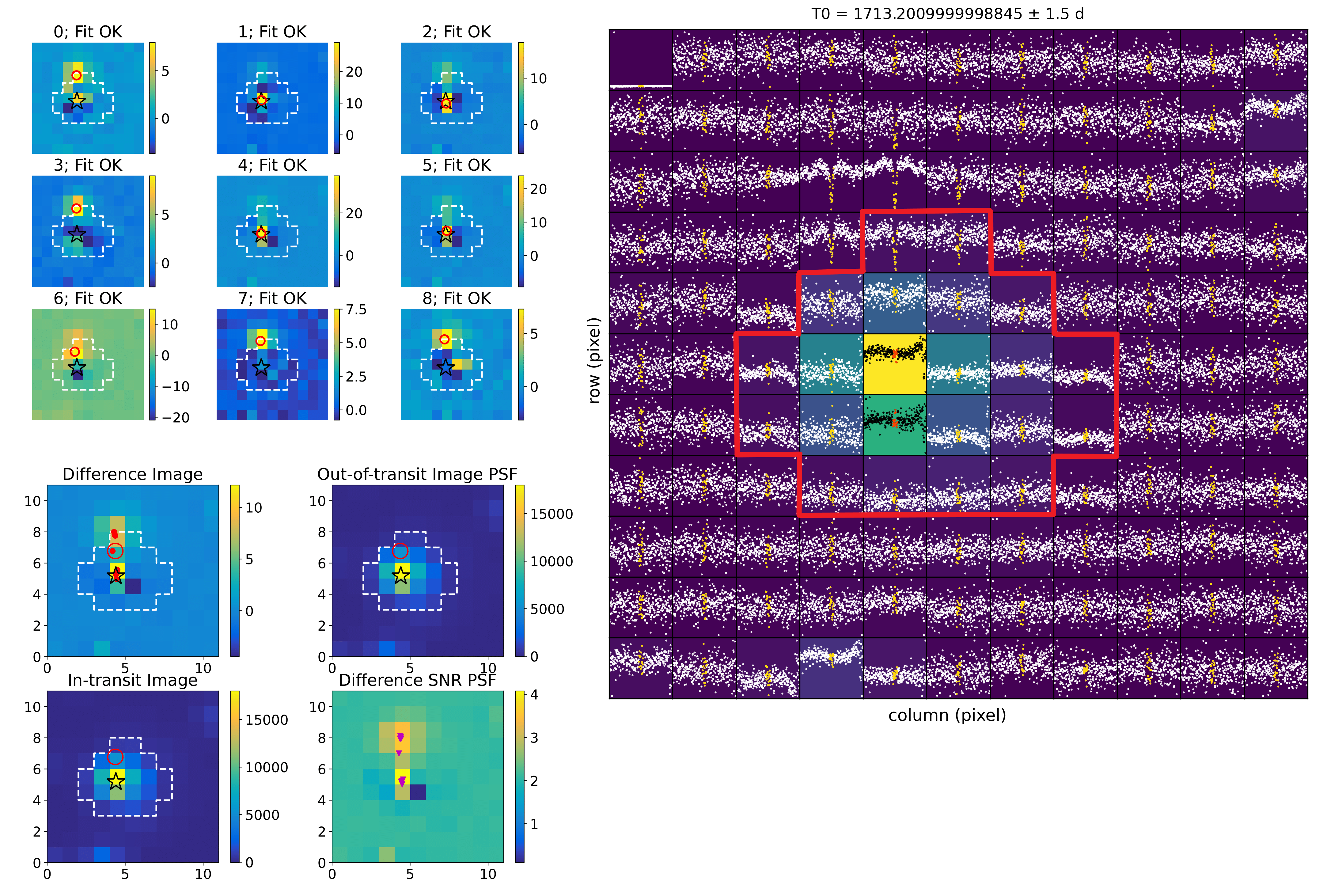}
    \caption{DAVE's photocenter analysis for TIC 256886630 (upper and lower left panels) and the corresponding PLL figure (right panel). The upper left panels shows difference images and the corresponding centroid measurements for $9$ transits detected in the TESS light curve observed at 2-min cadence in the Sector 15. Most of the difference images show a complex pattern instead of a single bright spot on an otherwise dark background. The corresponding photocenter measurements alternate between two distinct locations -- one near the target star and another few pixels above it. This makes interpreting the results from the photocenter module highly challenging. The PLL analysis on the right shows the first detected transit at $1713.20$ TJD. Clear eclipses are seen in several pixels away from the target, near the upper edge of the aperture mask (red contour). We see the same pattern for all the transits detected within sectors 15 and 16 where the TIC has been observed. This candidate is thus ruled out as a false positive because of CO.}
    \label{fig:plot_pixel_level}
\end{figure*}

Figure \ref{fig:plot_pixel_level} shows an example of such situation, highlighting how DAVE's measured photocenters for TIC 256886630 are unreliable due to the poor quality of many of the individual difference images (scenario (ii) above). Here, the PLL analysis immediately reveals that the true (and faint) source of the observed signal is near the edge of the aperture mask -- such that some of its signal does enter the aperture -- whereas the (much brighter) target shows no transit-like signal. As a result, this target has been ruled out as a false positive due to CO.

\subsection{Dispositions and comments}
According to the workflow described above, each of the \ntics{} TOIs presented here was thoroughly examined by at least three vetters, including at least one member of the core science team. The purpose of this workflow is twofold: to distribute the total workload over a large group of people, saving significant time and to also reduce the human bias that unavoidably affects inspection. Each vetter provides their evaluation (or \textit{disposition}) of the TOI under scrutiny, according to the following prescriptions: 

\begin{enumerate}
    \item if the TOI shows no anomalies at both the flux and the pixel level then the signal is ranked as a \textit{Planetary Candidate}, PC. We also classify the target as PC by default if any of the following cases are met:
    (a) the light curve has low SNR resulting in a very shallow dip and there are no indications for a centroid offset; 
    (b) the photocenter analysis generates unreliable centroids (UC) and the light curve does not show any obvious systematics; 
    (c) the phase-folded sector-by-sector light curve shows no apparent transit signal (NT) and there are no known nearby sources bright enough to produce the transit depth. We note that an NT flag is not unexpected since DAVE analyses individual sectors instead of multi-sector data. As a result, low SNR and/or long-period candidates may not have sufficient per-sector SNR for DAVE's tests.
    
    \item if the TOI does not pass the vetting procedure then the signal is ranked as a \textit{False Positive}, FP. A significant centroid offset (CO) represents one of the strongest clues for a FP scenario. A target is also classified as a FP when the phase-folded light curve exhibits a clear secondary eclipse (SS) or a significant OED. The latter is one of the most challenging features to distinguish as it is highly dependent on a quiet light curve;
    \item if \texttt{DAVE} generates a few red flags for a TOI but there are no clear indications of a false positive scenario, then the signal is ranked as a \textit{probable False Positive}, pFP. For example, a pFP may arise when the TESS light curve has a low SNR and at the same time we notice a potential secondary eclipse and/or the photocenter position seems to be slightly shifted towards a nearby field star. Long-period candidates are particularly difficult to analyze since the number of per-sector transits is small, and the measured photocenters might not be sufficient for a statistically-significant evaluation. Often, there are only one or two photocenter measurements. In cases like these, we flag the candidate as a pFP instead of FP even if the photocenter analysis indicates an offset.  
\end{enumerate}

\subsection{Automatic Disposition Generator}
\label{ADG}
In addition to the analysis described above, we also followed an additional procedure, that we named Automatic Disposition Generator (ADG), to automatically generate dispositions for TOIs based on the rankings of our vetters. For each TOI, we require dispositions from at least three vetters; the final disposition is determined by taking a weighted average of all vetters' dispositions. 
A critical step is to provide the ADG with a reliability indicator for each vetter via a \textit{user score} $\epsilon_{i}\in [0,1]$ to account for varying levels of expertise within our team. As the volunteers who contributed to this work are the same as those who contributed to the Paper I catalog, we used the latter's results to quantify the reliability of each vetter as follows. 

For each vetter, we constructed their own confusion matrix, as shown in Table \ref{tab:confusion_matrix}, using the final group dispositions of Paper I as our knowledge base.
\begin{table}
\centering
\begin{tabular}{|l|ll|}
\hline
                              & \multicolumn{2}{c|}{Actual label}  \\ \hline
\multirow{2}{*}{Vetter label} & \multicolumn{1}{l|}{T(PC)} & F(PC) \\ \cline{2-3} 
                              & \multicolumn{1}{l|}{F(FP)} & T(FP) \\ \hline
\end{tabular}
\caption{Confusion matrix of a single vetter. T(PC) represents the number of true PCs, T(FP) is the number of true FPs while F(PC) and F(FP) represent the number of TOIs that were incorrectly classified as PC and FP, respectively.}
\label{tab:confusion_matrix}
\end{table}

In Paper I, the true PCs accounted for $\sim 71\%$ of the total catalog over a total of $N_{\text{TOT}}=999$ targets. To account for the unbalanced nature of the knowledge base sample, we used the \textit{weighted average precision} as the metric to assess each vetter's level of reliability. Assume the $i$-th vetter ranked a certain number of targets in Paper I obtaining $T_i(PC)$ number of correctly identified PCs, $T_i(FP)$ number of correctly identified FPs, $F_i(PC)$ number of incorrectly identified PCs and $F_i(FP)$ number of incorrectly classified FPs, then their score $\epsilon_i$ will be given by the following 
\begin{equation}
    \epsilon_i=\dfrac{N_\text{PC}}{N_\text{TOT}}\left(\dfrac{T_i(PC)}{T_i(PC)+F_i(PC)}\right)+\dfrac{\widetilde{N}_\text{FP}}{N_\text{TOT}}\left(\dfrac{T_i(FP)}{T_i(FP)+F_i(FP)}\right) \, ,
\end{equation}
where $N_\text{PC}=709$ is the number of PCs in the catalog of Paper I while $\widetilde{N}_\text{FP}=290$ represents the number of both FPs and pFPs within the same catalog.  
Certainly, not all vetters have given the same number of dispositions, which may result in a non-uniform efficiency computation, but we ignore this as first-order approximation.

To calculate the weighted average of the overall disposition, we first convert labels into numbers, using the following convention:
\begin{equation}
	\text{PC}\equiv (1,0,0)\quad\quad\text{pFP}\equiv (0,1,0)\quad\quad\text{FP}\equiv (0,0,1) \, .
\end{equation}
Hence, we define the overall disposition as the vector $\vec{D}$ determined by the average of given dispositions weighted over the fidelity of vetters,
\begin{equation}
\centering
	\vec{D}\equiv (D_0,D_1,D_2)=\dfrac{1}{W}\left(\sum_{i=1}^{N_\text{PC}}\epsilon_\text{i},\sum_{j=1}^{N_\text{pFP}}\epsilon_\text{j},\sum_{k=1}^{N_\text{FP}}\epsilon_\text{k}\right),
	\label{overallD}
\end{equation}
where $N_\text{PC}$, $N_\text{pFP}$, $N_\text{FP}$ are the number of vetters who voted for PC, pFP and FP scenario respectively while $W\equiv\sum_{\ell=1}^{N}\epsilon_\ell$.

The final Paper Disposition is given using the following prescription:
\begin{equation}
	\begin{cases}
		\text{if}\quad\text{max}(\vec{D})=D_{0}\quad\Rightarrow\quad \text{Paper Disposition: PC}\\
		\text{if}\quad\text{max}(\vec{D})=D_{1}\quad\Rightarrow\quad \text{Paper Disposition: pFP}\\
		\text{if}\quad\text{max}(\vec{D})=D_{2}\quad\Rightarrow\quad \text{Paper Disposition: FP}\\
	\end{cases}
\end{equation}
In Table \ref{tab:scores} we reported the scores of each superuser who contributed to this work.
\begin{table}
\centering
\begin{tabular}{cc|cc|cc}
\hline
         & $\epsilon$ &          & $\epsilon$ &          & $\epsilon$ \\ \hline
Vetter 1 & 0.84       & Vetter 4 & 0.82       & Vetter 7 & 0.83       \\ \hline
Vetter 2 & 0.77       & Vetter 5 & 0.78       & Vetter 8 & 0.89       \\ \hline
Vetter 3 & 0.97       & Vetter 6 & 0.73       & Vetter 9 & 0.67       \\ \hline
\end{tabular}
\caption{User score $\epsilon$ for each superuserof this work.}
\label{tab:scores}
\end{table}
ADG not only drastically reduces the time required to generate a uniformly vetted catalog but it also allows for the reduction of human bias via a rigorous scientific approach. In this regard ADG captures the ultimate essence of a Citizen Science Project.

\section{Planet Patrol}
\label{sec:planetpatrol}

\cite{7106413} estimated that, on average and across all Zooniverse projects, citizen scientists inspected volumes of data equivalent to 34 years of full-time work by a single expert. For example, volunteers have discovered $41$ new long-period (Long-P) planet candidates in the Kepler database \citep{2015ApJ...815..127W} within the Planet Hunter TESS project \citep{2012MNRAS.419.2900F}. In three years citizen scientists involved in the PHT project helped the scientific team to discover hundreds of new planet candidates \citep{2022BAAS...54e.414E} along with a large number of eclipsing binary systems \citep{2021MNRAS.501.4669E}, including a hierarchical triple star system \citep{2022MNRAS.511.4710E}. Moreover, projects like the Visual Survey Group \citep{2022PASP..134g4401K} contributed to $69$ peer-reviewed papers mainly focusing on exoplanets, multistellar systems and unusual variable stars.  

The Planet Patrol project was officially launched on the 29th of September 2020 by the Zooniverse platform. The first stage of the project was aimed at improving the reliability of \texttt{DAVE}'s photocenter analysis by asking the trained users to evaluate the quality of the difference images generated by the \texttt{centroid} module. All users became acquainted with the workflow throughout brief vetting tutorials and F.A.Q. as well as numerous examples of false positives. 
More than $5,600$ volunteers examined $\sim 400,000$ difference images in just one month, achieving $95\%$ of accuracy using as a knowledge base $198$ classifications given by the science core team. 

After removing the difference images flagged as poor by the volunteers from \texttt{DAVE}'s analysis, the photocenter uncertainty decreased by up to $\sim 30\%$ for the majority of the candidates \citep{2022PASP..134d4401K}. 
After the completion of the first stage project (November 2020), many eager volunteers (superuser) expressed an interest in getting further involved in the vetting work. The superusers played a fundamental role in creating our first TT9 Catalog and repeated the feat by vetting the \ntics{} TESS candidates and assisting the core science team in producing the catalog presented here.

\subsection{Citizen scientists at work}

The main key to success of a citizen scientist project is having constant interaction between the science core team and the superusers. Hence, we hold live weekly meetings where we discuss the progress of the project and provide superusers the opportunity to discuss any difficulties they may have encountered throughout their task. 
Because our team is made of people from around the world, one of the superusers, HADL, recorded all meetings and posted them on a dedicated YouTube channel. These recordings (currently private) are useful for people not able to attend the specific meeting, and also provide a valuable resource for newcomers.

Citizen Science has taught us that volunteers can not only offer invaluable assistance in the specific scientific task, but also, due to their diverse expertise, they could provide the scientific community important new ideas, resources and tools. For example, as the Google Sheets we used to keep track of our vetting process grew in content and complexity, it became difficult to find, create, analyze, and distribute the user dispositions and comments. To address this issue, one of us (RS) developed a custom vetting portal, {\it Exogram} (\url{https://exogram.vercel.app}), specifically designed to streamline the vetting process and facilitate group discussions. {\it Exogram} is hosted by Vercel, the database and authentication is handled by Firebase, and parts of the backend logic is written in Python.

{\it Exogram's} homepage provides a user-friendly and intuitive interface that highlights targets that still need to be vetted by the user. It also allows the user to update their dispositions, search for TICs with specific dispositions or comments, and keep track of the overall vetting progress by all users. 
The website directly links each target to the \texttt{DAVE}-generated PDFs containing the vetting results and diagnostics stored on Google Drive. When creating dispositions, {\it Exogram} limits the user to three disposition options: False Positive (FP), Planet Candidate (PC), and Potential False Positive (pFP). The user comments section accepts both a pre-defined list of machine-readable vetting acronyms (e.g. ``CO'' for ``Centroid Offset'') and free text.  

\begin{figure*}
    \centering
    \includegraphics[width=0.85\textwidth]{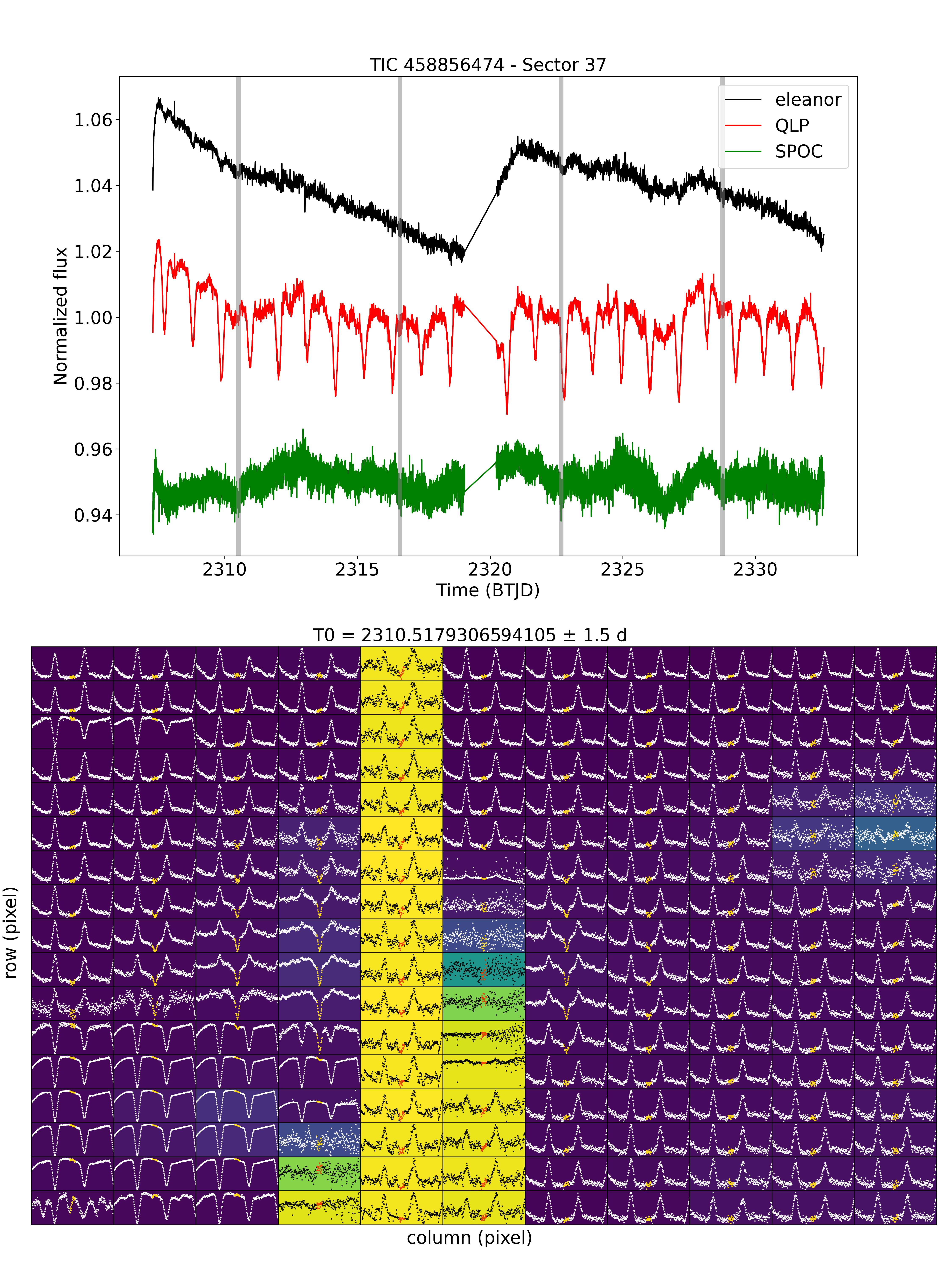}
    \caption{TIC 458856474.01 is a planet candidate orbiting its host star every $6.08$ days. In the upper panel we show the light curves of TIC 458856474 observed by TESS in sector 37 generated by \texttt{eleanor} (black) and QLP (red). The grey-shaded bars represent each transit within the sector. We also show the pre-processed SPOC light curve (green) for completeness. The light curve generated by \texttt{eleanor} is completely different from that of QLP. The latter is compatible with a prominent $\approx 1-$day eclipsing binary and a transiting object with a period of $6$ days. In the lower panel the PLL analysis for the first transit in sector 37 manages to solve the conflict. In fact, it clearly shows that the $1-$day eclipsing binary signal originates from a nearby pixel within the aperture mask used by QLP to extract its own custom light curve.}
    \label{fig:qlp_eleanor}
\end{figure*}

In addition, {\it Exogram} allows the user to interactively inspect and manipulate the target's light curve. The website downloads all available QLP data on demand from MAST, displays the corresponding normalized flux, centroid motion, and background flux for one or multiple TICs, and highlights the recorded momentum dumps. This allows the vetter an additional layer of scrutiny beyond that provided by \texttt{DAVE}, a complementary comparison between light curves produced by two different pipelines (\texttt{eleanor} vs QLP), and enables the user to explore and examine in detail the light curves of nearby targets. Figure \ref{fig:qlp_eleanor} represents one of the more interesting targets within our catalog for which the QLP's light curve is completely different from that generated by \texttt{eleanor}.

\section{The catalog}
\label{sec:catalog}
The \ntics{} candidates analyzed in this work were drawn from the candidates provided by the ExoFOP TESS archive in the fall of 2020. They were selected by TIC number and do not overlap with our first TT9 catalog. Once each TOI had at least 3 dispositions, we ran ADG on the whole catalog. 
It generated $752$ signals as PCs, $142$ as FPs and 105 as pFPs. Thus, overall approximately one in three planet candidates is a false positive or a potential false positive, a rate similar to that of Paper I. The most common comments within our catalog are ``FSCP'' and ``FSOP'' which occurred $628$ and $481$ times respectively. This is expected, given that TESS targets are often contaminated by nearby background and/or foreground sources. We note that we only use these two flags as an extra layer of scrutiny -- they are not sufficient to mark a candidate as a false positive.

\subsection{Planet candidates}
\label{sec:planet_candidates}

Within our catalog 752 TOIs passed all \texttt{DAVE} tests and human inspections as planet candidates. In this sample there are $117$ objects that have already been confirmed within the TESS scientific community or previously discovered by other exoplanetary surveys.
Twelve of the $752$ PCs can be regarded as bona-fide, high-quality candidates, as they passed the \texttt{DAVE} test showing a clear box-shaped transit and high-significance on-target centroid measurements. In Table \ref{tab:12strongPC} we summarize the main properties of these $12$ likely genuine planets. None of these $12$ candidates has been confirmed by follow-up observations yet.

\begin{table*}
\begin{threeparttable}
\begin{tabular}{lllllllllll}
\hline
TIC ID & TOI    & TESS sectors   & \begin{tabular}[c]{@{}l@{}}Epoch \\ (BTJD\tnote{†}\;)\end{tabular} & \begin{tabular}[c]{@{}l@{}}Period \\ (d)\end{tabular} & \begin{tabular}[c]{@{}l@{}}Duration \\ (h)\end{tabular} & \begin{tabular}[c]{@{}l@{}}Depth\\ (ppm)\end{tabular} & \begin{tabular}[c]{@{}l@{}}$R_p$ \\ $(R_J)$\end{tabular} & \begin{tabular}[c]{@{}l@{}}$R_S$\\ $(R_\odot)$\end{tabular} & $\text{TESS}_{\text{mag}}$ & Comment             \\ \hline
375542276 & 1163.01 & 14,40,41                                                                    & 2459441.80                                              & 3.08                                                  & 2.27                                                    & 4800                                                  & -                                                        & -                                                           & 9.42                & FSCP                \\
417948359 & 1272.01 & 15,16,22,49                                                                 & 2459661.40                                              & 3.32                                                  & 1.55                                                    & 2770                                                  & 0.38                                                     & 0.81                                                        & 11.02               & FSCP                \\
439456714 & 277.01 & 3,30                                                                        & 2458385.03                                              & 3.99                                                  & 2.04                                                    & 5801                                                  & 0.39                                                     & 0.52                                                        & 11.73               & FSCP                \\
348770361 & 161.01 & \begin{tabular}[c]{@{}l@{}}1,12,13,\\ 27,28,39\end{tabular}                 & 2459388.79                                              & 2.77                                                  & 4.62                                                    & 3234                                                  & 0.57                                                     & 0.98                                                        & 11.50               & FSOP                \\
468148930 & 1086.01 & 13,27                                                                       & 2458655.29                                              & 3.72                                                  & 5.59                                                    & 4872                                                  & 0.74                                                     & 1.15                                                        & 12.23               & FSOP, Vshape                \\
459969957 & 1274.01 & \begin{tabular}[c]{@{}l@{}}14,15,17-26,\\ 40,41,47-52,\\ 54,55\end{tabular} & 2459736.34                                              & 19.32                                                 & 4.31                                                    & 13910                                                 & 0.82                                                     & 0.80                                                        & 11.90               & HPMS, Long-P                \\
272758199 & 1845.01 & \begin{tabular}[c]{@{}l@{}}14-17,19-23,\\ 26,40,41,\\ 47,49,50\end{tabular} & 2459666.13                                              & 3.66                                                  & 2.52                                                    & 19640                                                 & 1.23                                                     & 0.94                                                        & 12.89               & Short-P\tnote{§}, pVshape                 \\ 
147456499 & 2659.01 & 3,30                                                                        & 2459140.20                                              & 1.25                                                  & 1.79                                                    & 22260                                                 & 1.29                                                     & 0.90                                                        & 12.73               & Short-P, pVshape\\
441797803 & 1302.01 & \begin{tabular}[c]{@{}l@{}}14-19,21,\\ 22,24,25,\\ 41,47-52\end{tabular}    & 2459738.40                                              & 5.67                                                  & 3.67                                                    & 9780                                                  & 1.29                                                     & 1.47                                                        & 10.67               & FSOP, pVshape \\
394561119 & 1107.01 & 11-13,38,39                                                                 & 2459385.01                                              & 4.08                                                  & 4.80                                                    & 5789                                                  & 1.30                                                     & 1.75                                                        & 10.01               & LCMOD, Vshape, FSOP \\
252616865 & 1482.01 & 16,17                                                                       & 2458741.29                                              & 5.71                                                  & 4.74                                                    & 7600                                                  & 1.39                                                     & 1.79                                                        & 10.07               & FSOP                \\
289539327 & 1186.01 & \begin{tabular}[c]{@{}l@{}}14-26,40,\\ 41,47-55\end{tabular}                & 2459817.02                                              & 11.21                                                 & 7.88                                                    & 8990                                                  & 1.76                                                     & 1.96                                                        & 9.92                & FSCP, pVshape                \\
\hline
\end{tabular}
\begin{tablenotes}
\item[†] Baricentric Truncated Julian Date
\item[§] Short period candidate
\end{tablenotes}
\end{threeparttable}
\caption{List of the $12$ most promising planet candidates in our work. For each TOI we report the TIC and TOI identifiers, the \texttt{DAVE} input parameters, the radius of the transiting object $R_p$, the stellar radius $R_*$ along with its TESS magnitude and the final comments provided by the vetters.}
\label{tab:12strongPC}

\end{table*}

Apart from the ``FSCP'' comments that are quite spread all over the catalog, the most common comments for our PCs are ``LowSNR'' ($270$ times), ``UC'' ($247$ times), ``LCMOD'' ($201$ times) and ``Vshape'' ($147$ times). 
The first two comments are strongly correlated because \texttt{DAVE} often generates unreliable centroids for signals with low SNR, thus making the classification challenging. 
As already discussed, in these cases we automatically flag the target as PCs. The third most notable comment can either be caused by the inherent modulations of the targets under scrutiny or from the sources that contaminate the extracted light curve. Strong light curve modulations can also completely hide shallow transits that could be identified after careful detrending. 
Finally, the flag for V-shaped transit is not a conclusive evidence to support a false positive scenario. It only indicates that the two objects orbiting a common center of mass have comparable sizes. Although this happens more frequently for a binary star system, we can not rule out a giant planet transiting its host star with a non-zero impact parameter \citep[e.g., ][]{2011A&A...526A.130S}.

\subsection{False positives}

Our analysis classified 142 candidates as FP. Of these, we ruled out 118 targets as false positives due to a clear ``CO''. While nearly $40\%$ of false positives in Paper I was flagged as ``CO'', in this work the rate has increased to approximately $83\%$. The PLL analysis was likely essential for some targets that otherwise would have been flagged as PC because of poor centroid measurements. Furthermore, we believe that the observed percentage increase in the "CO" flag is also due to the volunteers' skill improvement after two years of training.

 The second most frequent false positive indicator is the presence of a significant secondary eclipse (``SS'', 33 targets), followed by Odd-Even Difference (``OED'', 34 targets). Both of these flags are often accompanied with a ``Vshape'' comment. All OED targets have been inspected for prominent modulations of the light curve. 

We note that we vet all the TOIs presented here regardless of their current disposition on ExoFOP as done in Paper I. In particular, in Paper I six confirmed planets were classified as FP due to a significant secondary eclipse at mid-transit. In this work, out of $142$ targets, we labelled as FP TIC 427761355.01 and TIC 386259537.01 that have have confirmed as bonafide planets by follow-up observations. 
The \textit{Modelshift} of TIC 427761355.01 (or TOI-1518 b) shows a V-shaped transit with a SS exactly at half period. At this level of significance we cannot distinguish a secondary eclipse from a planetary occultation, thus for consistency with our workflow we flag it as a ``FP''. 
We also labelled WASP-169 b as a FP since its \textit{centroids} module depicts a clear and reliable offset of the light photocenter at the time of transit. After inspecting this target with PLL, we discovered that there is a deeper transiting feature in the nearby pixel on the same period of WASP-169 b, which causes the overall centroid to shift.

\subsection{Potential false positives}

We labelled 105 TOIs in our catalog as pFPs. Our concerns and difficulties towards these targets are reflected in the most notable comment, ``potential-CO'' ($61$ times). The prefix ``potential'' qualitatively indicates that we are not fully convinced there is a significant photocenter offset due to ``LowSNR'' ($59$ times) or prominent ``LCMOD'' ($40$ times) which complicated the centroid measurements. It often happens that among many unreliable centroid measurements there are a handful that show a hint for a CO. In these cases, we could not eliminate our concerns with the PLL analysis either. However, it did help us identify as pFP three targets which were previously classified as PC due to UC.
The light curves of these targets usually do not show a clear transit ('LowSNR', $59$ times) leading to $26$ cases for which a potential secondary eclipse has been observed as well as $20$ cases where an OED might be statistical significant. 

\subsection{Individual targets of interest}

\begin{figure}
    \centering
    \includegraphics[width=0.48\textwidth]{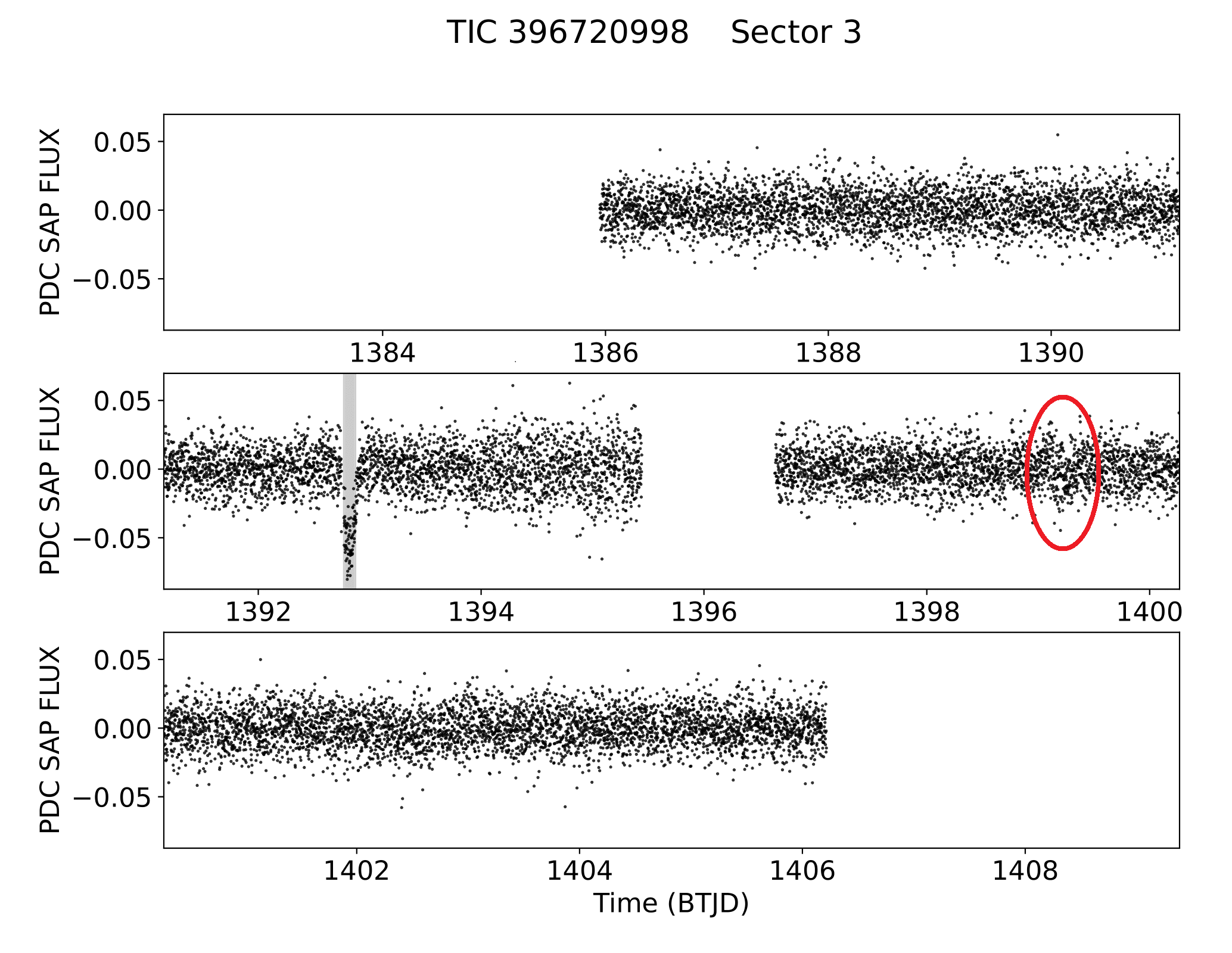}
    \caption{The light curve of planet candidate TIC 396720998.01 as observed by TESS in sector 3. The grey-shaded bar highlights the transit as detected by the SPOC pipeline. In addition we noted a potential secondary feature at $\approx 1399$ BTJD. We found a correspondence in the ExoFOP archive which flagged this signal as the candidate TIC 396720998.02. Its reported orbital period is $\approx 777$ days that may be an upper limit due to the lack of observations between sectors 5 and 30.}
    \label{fig:hot_white}
\end{figure}

One of the most intriguing and worth noting planet candidate within our catalog is TIC 396720998.01, a sub-Jovian ($R_p\approx0.35 \, R_J$) object orbiting a white dwarf ($R_* \approx 0.15 \, R_\odot, M_*\approx0.5M_\odot$ and $T_*\sim 50,000K$), according to the Tess Input Catalog. It has been observed by TESS in sectors 3,4,5,30,31 and 32. We also found additional transit-like features ($\approx 6000$ ppm) that may suggest a multiplanet system around this hot white dwarf as shown in Fig. \ref{fig:hot_white}. 
We flagged a V-shaped transit potentially due to the small size of the host star ($\approx 0.15 \, R_\odot$). This system could represent a perfect target to shed light on the evolution of a planetary system around Sun-like stars during the last stages of their evolution.

Among the TOIs listed in our catalog, we also kept track for planet candidates orbiting within the so-called habitable zone of their host stars. 
For each given star with known radius $R_*$ and mass $M_*$ we calculated the inner and outer edges of its so called habitable zone as defined by \cite{2013ApJ...765..131K}. As to the inner edge we considered the runaway greenhouse at which the oceans evaporate entirely, while the outer edge was calculated considering the maximum greenhouse provided by a CO$_2$ atmosphere. We found two planet candidates that orbit the habitable zone of the stars TIC 271971130 and TIC 360156606.

TIC 271971130.01 is a planet candidate with $R\approx 1.6 R_\oplus$ and $P\approx 19.3$ days detected by the SPOC pipeline. TESS observed the target in sectors 1-13, 27, 29-37, and 39 at cadences of 2, 10 and 30 minutes. This TOI is marked in our catalog as a LowSNR candidate; in some sectors it is quite challenging to see the transits. It is a faint star (TESS$_\text{mag}=13.5$) for which we also flagged ``FSCP'' and ``FSOP'''. 
Hence, the light curve is contaminated by nearby fainter sources ($<15$ TESS$_\text{mag}$) within the aperture mask and the same pixel. 
As discussed in Sect. \ref{sec:method}, in cases like this we consider the candidate as a PC by default. According to the TESS Input Catalog stellar parameters the host star is a red M dwarf with $T_* \approx 3187 \, K, R_*\approx 0.22 \, R_\odot$ and $M_*\approx  0.20 \, M_\odot$. The candidate planet lies very close to the inner edge of the habitable zone of its star. 

TIC 360156606.01 is a planet candidate with $R\approx 9 R_\oplus$ and $P\approx 27.36397$ days. Its host star has been observed by TESS in sectors 11 and 12 at a cadence of both 2 and 30 minutes, and in sector 38 at 20 seconds, 2 and 10 minutes cadences. The planet candidate is marked in our catalog as a LowSNR signal. Its light curve shows prominent modulations which make the Modelshift analysis difficult. These modulations may originate from brighter sources that fall within the aperture mask. However the detected transit is above the noise and quite clear. As discussed above, due to its long orbital period the photocenter test from DAVE is inconclusive. According to the TESS Input Catalog Stellar Parameters the host star is a red M dwarf with $T_*\approx 3055 K, R_*\approx 0.446R_\odot$ and $M_*\approx 0.43M_\odot$. This candidate has been recently confirmed by \cite{2022AJ....163..156M} as TOI-1227 b within the TESS Follow-up Observing Program Working Group.

\subsection{Comparison to dispositions based on Machine Learning}
As we mentioned in the Introduction, Machine-learning based pipelines are also effective in providing dispositions 
for a large sample of TOIs. To date, there are two main algorithms based on deep learning that have been explicitly tested on TESS data: the \texttt{ASTRONET} versions described in \citet{Yu2019} and in \citet{Tey2023}; and \texttt{Exominer} \citep[][see their Section 10]{Valizadeganetal}. 
\citet{Yu2019} describe five different networks, with different tasks, ranging from the ``triage'' model that only works on light curves and removes false positive signal produced by instrumental artifacts, to vetting models that can also take into account analyses centroids' positions and additional information. 
Their best vetting model achieves an average precision \footnote{In the context of Machine-Learning, "Precision" indicates the ratio between true positives over all predicted positives, "recall" indicates all true positives over all data with positive labels, and "accuracy" indicates the number of true positives and true negatives over the total number of samples.} of 69.3\% and an accuracy of 97.8\%. 
Their algorithm has recently been improved by \citet{Tey2023}, reaching a 99.6\% recall at a precision of 75.7\%.
On the other hand, \texttt{Exominer} makes use of the unique elements of Kepler SOC/TESS SPOC data validation summary report in their original format. \texttt{Exominer} reaches a 88\% precision at the recall value of 73\% on TESS data. 
Following a similar approach to that of \texttt{ASTRONET}, \citet[]{Fiscale21} and \citet[][hereinafter F23]{Fiscale22}, presented a deep-learning method to obtain dispositions from TESS data. 
By working only on the light curves, the model described in F23 achieves a precision of 87\% at recall value of 81\%.
Note that applying Neural Network models described in the literature to an arbitrary dataset is not straightforward: it requires additional work, even when the code is publicly available (as in the case of e.g. \texttt{ASTRONET}), including preforming the training from scratch \citep[see e.g. the discussion in][]{Visser2022}. Besides reaching such good performances, the F23 model has the additional advantage of being developed within our research group and therefore can be immediately applied to the catalogs discussed in this work.
We first tested the F23 network on the Paper I catalog, in order to use the algorithm described in Sect.~\ref{ADG} to estimate its score, that is 0.73, hence similar or better than the one of a third of the superusers. 

Therefore, we can compare the independent dispositions obtained by the neural network with the catalog obtained by exploiting citizen science, in order to check for consistency. 

\begin{figure}
    \centering
    \includegraphics[width=0.5\textwidth]{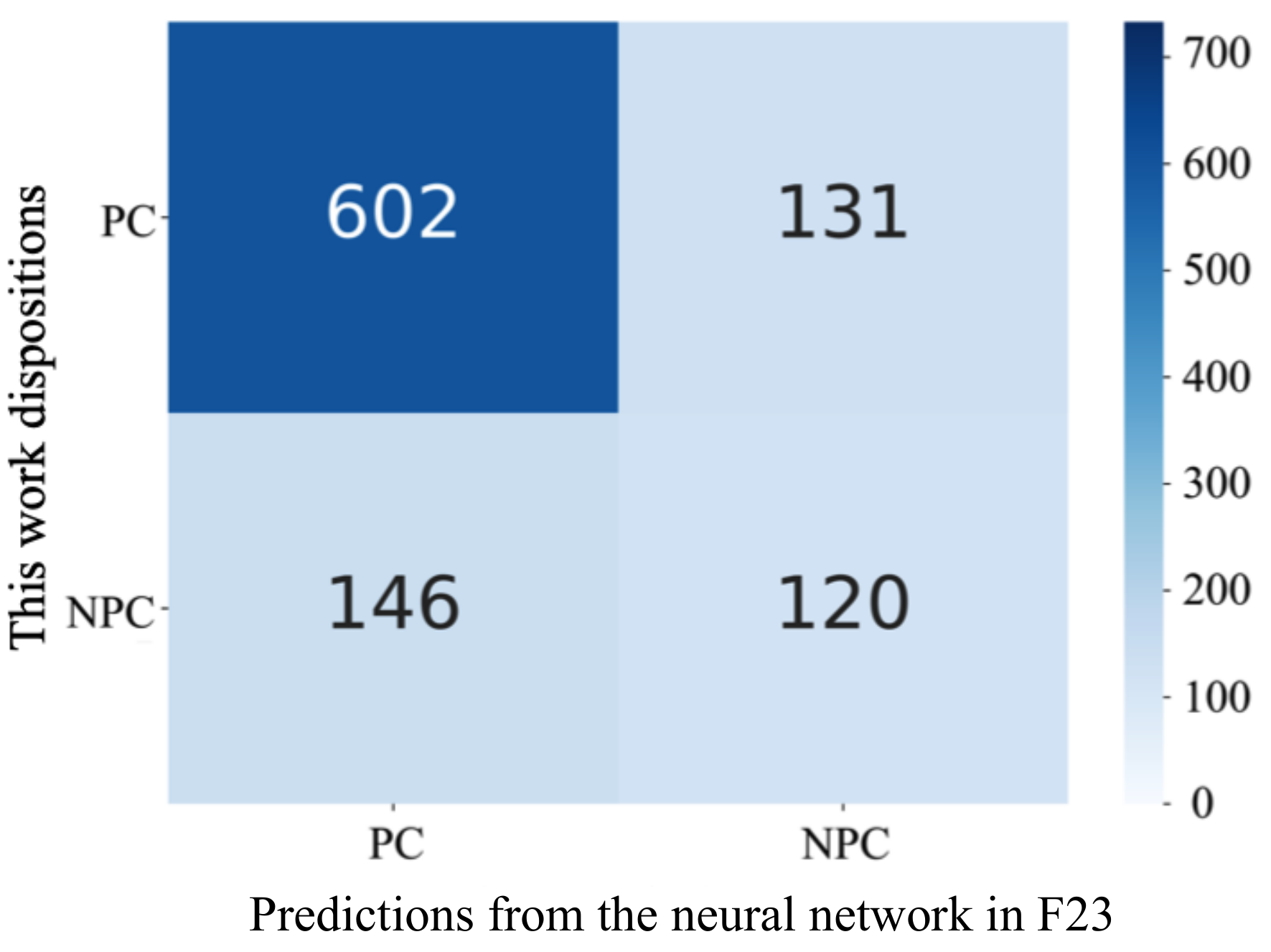}
    \caption{Confusion matrix of the neural network we applied on the \ntics{} TOIs. From top-left to bottom-right, four outcomes produced by this network are shown: true positives (TP), false positives (FP), false negatives (FN) and true negatives (TN). On the basis of this confusion matrix, the F23 model achieves 80\% precision and 82\% recall, with an accuracy of 72\%.}
    \label{fig:confmat_cnn}
\end{figure}

We show this comparison in the form of the confusion matrix in Fig. \ref{fig:confmat_cnn}, where we consider our catalog's outcome as ground truth. Hence, true positives (TP), true negatives (TN), false positives (FP) and false negatives (FN) are computed with respect to our dispositions. 
Specifically, the TP and TN represent the fractions of TOIs classified by both our team and the network as PC and not PC respectively. The FP indicates the fraction of TOIs we labelled as not planet candidate (i.e. we classified as FP or pFP) while the network predicts as PC, and FN indicates the number of TOIs that are indicated as PC in our catalog but are not identified by the network.

Furthermore, over half of the TOIs mislabelled as not planet by the network are flagged as LowSNR targets in this work, with half of these targets not showing any visible transit. In these cases, we decided to be conservative and pass the signal as a candidate if there are no other issue. The network however is trained on datasets where similar objects are not labelled as PC, hence it cannot provide the same disposition as us.
Summarising, we find that machine-learning approaches are promising, but they still need to be complemented with the study of the ancillary data available (such us the photocenter position) in order to provide final dispositions and validation, especially in the case of low SNR light curves. 

\section{Discussion}
\label{sec:discussion}
\begin{figure*}
    \centering
    \includegraphics[width=\textwidth]{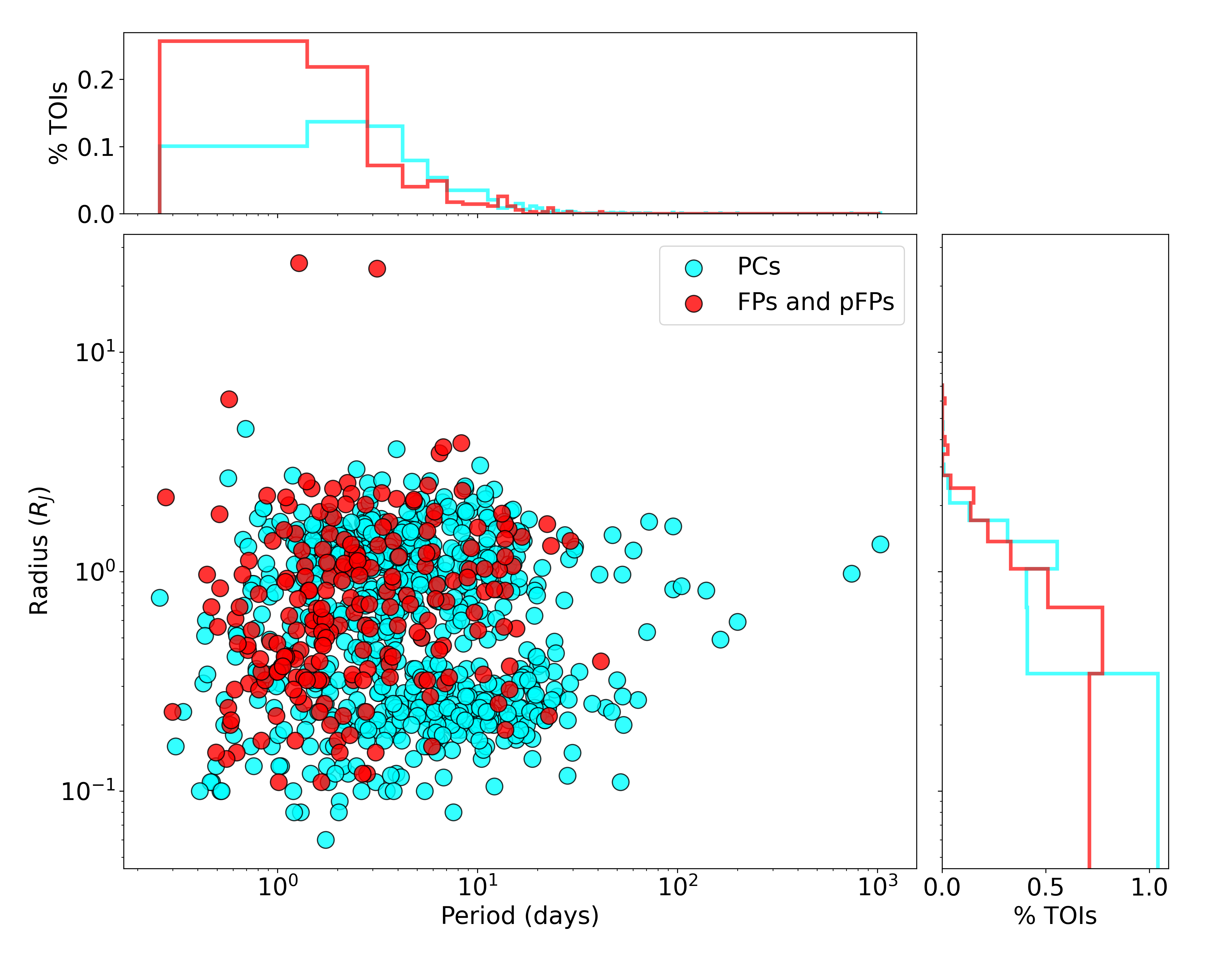}
    \caption{The distribution of the \ntics{} targets of the catalog within the $(P, R)$ diagram. The cyan dots represent the bona-fide planet candidates while the red dots represent the false positives and potential false positives.}
    \label{fig:PR_distribution}
\end{figure*}

In Fig. \ref{fig:PR_distribution} we show the distribution of the \ntics{} TOIs within the $(P, R)$ plane. The figure highlights the high rate of false positives at short periods and large planetary radius. A potential explanation for this result may indicate that the majority of false positive scenarios originate from close eclipsing binary systems. We also emphasize that our procedure automatically classifies all long period candidates (> 50 days) as PC because these objects have insufficient per-sector photocenter measurements and are usually flagged as 'LowSNR' candidate.

We also applied a two-sample Kolmogorov-Smirnov test to the orbital period distributions of PCs and pFPs-FPs obtaining a $p$-value less than 0.05. We repeated the test for the radii distribution between PCs and pFPs-FPs obtaining the same result. This suggests that the two samples come from different distributions within the fixed level of confidence. 
These trends are in agreement with those obtained in Paper I; in particular we did not find any statistical deviations in the $P$ and $R$ distributions between the same classes of the two catalogues. This was expected since the methodology underlying both catalogues is practically the same. Hence we merged the two catalogues in one sample containing $1998$ uniformly-vetted TESS candidates. We performed a statistical analysis of this sample by taking into account the statistical correlation between the orbital period and the planetary radius in the planet rate occurrence \citep{Hsuetal}. Hereafter, we will use (p)FPs when we refer to the both FPs and pFPs contained in the sample. 

\begin{figure*}
    \centering
    \includegraphics[width=\textwidth]{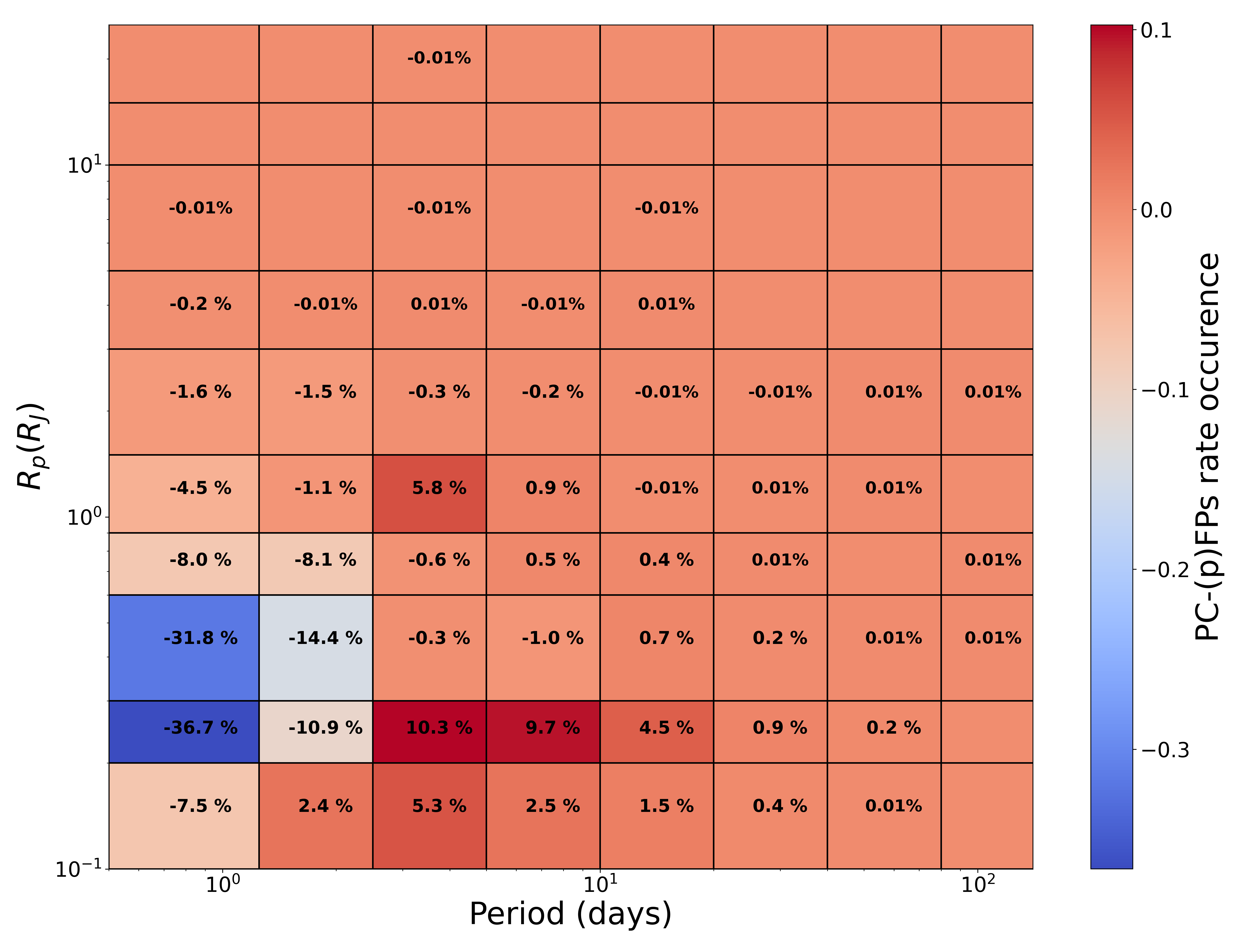}
    \caption{Period-radius occurrence rates of the difference between planet candidates and false positives (including the potential false positives) for the whole sample of 1998 targets investigated in this work and in Paper I. The numerical values of the occurrence rates are expressed as percentages. We note that the bin sizes are not uniform. Blank cells are those that contain neither PCs nor (p)FPs.}
    \label{fig:FPs_rateoccurrence}
\end{figure*}

In Figure \ref{fig:FPs_rateoccurrence} we show the difference between the occurrences of PCs and (p)FPs within the $(P, R)$ diagram.
When considering the orbital period and the planetary radius at the same time, we observe that the (p)FPs still outnumber the PCs at short period ($P \lesssim 4$ days) but the dependence on the planetary radius is more complex. In particular for $P\leq 4$ days, the PCs under-dense cells form a triangular shape region that overlaps the so-called Hot Neptune Desert. Demographic studies revealed a scarcity of discovered exoplanets within this region \citep{2011ApJ...727L..44S,2013ApJ...763...12B}. Hence, our analysis suggests that most of the planet candidate signals falling within the Hot Neptune Desert are consistent with false positives. This is also consistent with the results of \cite{Magliano2022} who classified a sample of Hot Neptune candidates using the same methodology. In particular, in their sample of TESS candidates with $P\leq 4$ days and $0.27\leq R_p\leq 0.44 R_{J}$, nearly $75\%$ of the investigated candidates were flagged as (p)FP. The rate occurrences obtained here could be also used as priors to develop a Bayesian pipeline aimed at vetting a batch of TESS candidates.

\section{Conclusions}
\label{sec:conclusions}

We presented our second catalog of \ntics{} uniformly-vetted transiting exoplanet candidates from TESS as part of the Planet Patrol citizen science project. We implemented new diagnostics within our workflow to help vetters scrutinizing the more challenging cases. We also introduced a more precise way of getting a final group classification based on vetter's reliability. We marked $752$ TOIs as planet candidates, of which $117$ are confirmed planets. We also identified $12$ planet candidates which passed all the vetting diagnostics placing themselves as high-priority targets to be confirmed.
$142$ TOIs have been classified as false positives mainly due to a clear offset in the measured photocenter and/or a significant secondary eclipse. To be consistent with our workflow we found out that $2$ targets labelled as false positives were true planets. Finally, $105$ TOIs were flagged as potential false positives due to a potential centroid offset or secondary eclipse dominated by light curve modulations and/or systematics.
Together with Paper I, this work creates a catalog of uniformly-vetted TOIs that can be further used to prioritize targets amenable for follow-up observations.
Additionally, the two catalogs can be utilized as a training set for machine learning efforts aimed at full automation of the vetting process. This catalog is provided to the scientific community in the same format as the Table \ref{tab:catalog}; the full table is available as supplementary material along with this manuscript. 
The files generated by \texttt{DAVE} are publicly-available on the Exogram platform and will be also made available on ExoFOP-TESS as part of the metadata associated with each TOI.

\begin{landscape}
\begin{table}
\begin{threeparttable}
\begin{tabular}{llllllllllll}
\hline
TIC ID     & TOI     & Disposition & Comments                             & Sector(s) & Period (d) & Duration (h) & Depth (\%) & $R_p(R_J)$ & $R_S(R_\odot)$ & $\text{TESS}_{\text{mag}}$ & $\Delta\text{TESS}_{\text{mag}}$ \\ \hline
279177746  & 1353.01 & FP          & LowSNR, CO, FSCP                      & 15,16     & 4.43126    & 1.72         & 0.18       & 0.78       & 1.87           & 10.2    & 6.9             \\
387259626  & 1455.01 & FP          & SS, TD                               & 15,16,17  & 3.62339    & 2.59         & 1.63       & 1.69       & 1.35           & 10.2    & 4.5             \\
255760319  & 814.01  & FP          & SS, CO                               & 6,8       & 7.21048    & 2.67         & 0.03       & 0.33       & 2.14           & 9.0     & 9.0             \\
407394748  & 1623.01 & FP          & Vshape, SS, FSCP                     & 25        & 1.50455    & 1.11         & 0.11       & 0.60        & 1.77           & 9.5     & 7.4             \\
256886630  & 1323.01 & FP          & CO                                   & 15,16     & 2.03905    & 2.26         & 0.04       & 0.57       & 2.68           & 8.3     & 8.4             \\
410528770  & 1502.01 & FP          & Vshape, CO, FSCP                     & 16,17     & 2.75347    & 3.46         & 0.29       & 2.02       & 2.41           & 10.5    & 6.4             \\
270238522  & 1514.01 & FP          & CO, LowSNR                            & 17,18     & 1.36991    & 4.16         & 0.12       & 1.06       & 3.64           & 7.6     & 7.3             \\
425163745  & 937.01  & FP          & OED, Vshape, CO                      & 5         & 0.27664    & 0.72         & 1.57       & 2.18       & 1.78           & 10.9    & 4.5             \\
425721385  & 1128.01 & FP          & CO, FSCP                             & 13        & 13.5499    & 1.80          & 0.12       & 0.56       & 1.63           & 9.7     & 7.3             \\
287474726  & 582.01  & FP          & FSCP, pSS, CO                        & 8         & 3.74237    & 2.48         & 0.06       & 0.95       & 4.24           & 9.4     & 8.0             \\
307734817  & 1808.01 & FP          & FSCP, CO                             & 22        & 2.1075     & 1.91         & 0.43       & 1.05       & 1.59           & 11.7    & 5.9             \\
356235833  & 2111.01 & FP          & CO, Vshape, pSS                      & 25        & 1.26959    & 0.98         & 0.07       & 0.27       & 0.95           & 9.2     & 8.0             \\
365639282  & 482.01  & FP          & Vshape, FSOP, CO                     & 6         & 10.8717    & 2.94         & 1.46       & 0.81       & 0.49           & 13.1    & 4.6             \\
386259537  & 1932.01 & FP          & CO, KP \tnote{†}, Vshape                       & 7,8       & 5.61264    & 6.24         & 0.47       & 1.53       & 2.28           & 11.3    & 5.8             \\
350332997  & 832.01  & PC          & FSOP, FSCP                           & 11        & 1.91693    & 1.73         & 0.25       & 0.56       & 1.07           & 11.8    & 6.5             \\
350743714  & 165.01  & PC          & FSCP                                 & 3         & 7.76178    & 3.49         & 0.44       & 1.23       & 1.87           & 9.8     & 5.9             \\
353782445  & 1664.01 & PC          & FSOP                                 & 18,19     & 11.83441   & 2.24         & 0.03       & 0.22       & 1.34           & 9.2     & 8.7             \\
256783784  & 1432.01 & PC          & LowSNR, HPMS, FSCP                   & 15,16     & 6.11219    & 2.19         & 0.06       & 0.19       & 0.91           & 9.5     & 8.2             \\
259377017  & 270.03  & PC          & LowSNR, FSOP                         & 3,4,5     & 3.360137   & 1.39         & 0.10        & 0.23       & 0.42           & 10.5    & 7.5             \\
269558487  & 855.01  & PC          & FSOP, Short-P                        & 3         & 1.8306     & 1.19         & 0.11       & 0.38       & 1.22           & 10.7    & 7.4             \\
271900960  & 389.01  & PC          & UC, LCMOD                            & 4         & 13.45913   & 3.43         & 0.26       & 0.69       & 1.38           & 8.3     & 6.5             \\
276128561  & 829.01  & PC          & FSCP, UC                             & 11        & 3.287693   & 2.26         & 0.17       & 0.40        & 1.08           & 10.6    & 6.9             \\
277634430  & 771.01  & PC          & LowSNR, UC, FSCP                     & 10        & 2.325931   & 1.18         & 0.34       & 0.59       & 1.00              & 12.1    & 6.2             \\
277683130  & 138.01  & PC          & Vshape                               & 1         & 6.198041   & 2.18         & 0.37       & 0.74       & 1.11           & 9.6     & 6.1             \\
278198753  & 936.01  & PC          & FSCP                                 & 12        & 7.942733   & 1.69         & 0.68       & 0.60       & 0.70            & 12.5    & 5.4             \\
278348461  & 1257.01 & PC          & FSCP, KP, UC              & 14        & 5.452752   & 3.53         & 0.75       & 1.24       & 1.57           & 9.9     & 5.3             \\
278866211  & 189.01  & PC          & Vshape, Short-P                      & 1,2       & 2.194097   & 1.96         & 0.57       & 1.32       & 1.12           & 10.3    & 5.6             \\
279425357  & 739.01  & PC          & FSCP                                 & 9         & 9.013951   & 1.32         & 0.54       & 0.84       & 1.16           & 11.6    & 5.7             \\
409520860  & 1547.01 & pFP         & pCO, LCMOD, FSCP, LowSNR, NT         & 17        & 0.71933    & 1.47         & 0.05       & 0.31       & 1.44           & 10.4    & 8.3             \\
285677945  & 1571.01 & pFP         & LowSNR, NT, FSOP, FSCP, pCO          & 18        & 1.2427     & 1.56         & 0.07       & 0.54       & 2.14           & 10.2    & 7.9             \\
259389219  & 915.01  & pFP         & Vshape, pOED, pSS, FSCP              & 3,4,5     & 2.32485    & 1.51         & 0.22       & 0.76       & 2.05           & 11.2    & 6.7             \\
375225453  & 1096.01 & pFP         & LCMOD, pCO                           & 10,11,12  & 0.92145    & 1.42         & 0.04       & 0.48       & 2.89           & 10.1    & 8.6             \\
268301217  & 1937.01 & pFP         & FSOP, shortP, Vshape, FSCP, pOED     & 7,9       & 0.94667    & 1.27         & 1.38       & 1.38       & 1.07           & 12.5    & 4.7             \\
197959526  & 2303.01 & pFP         & Vshape, pSS, pCO, FSCP, FSOP         & 28        & 0.7196     & 1.47         & 0.50        & 1.12     & 1.69        & 11.8    & 5.8             \\
290348382  & 1099.01 & pFP         & FSCP, LowSNR, pCO, NT                & 13        & 6.44049    & 1.85         & 0.79       & 0.44       & 0.53           & 9.7     & 5.3             \\
316937670  & 221.01 & pFP         & LowSNR, FSCP, FSOP, pCO, Short-P       & 1,2        & 0.62425   & 0.77         & 0.10       & 0.15        & 0.53              & 12.2    & 7.5             \\
467281353  & 1975.01 & pFP         & LCMOD, pOED, UC, Vshape, FSCP,   pEV & 10,11     & 2.82888    & 3.25         & 0.07       & 0.36       & 1.48           & 10.2    & 7.8             \\
470852531  & 1625.01 & pFP         & NT, pCO, FSCP, Short-P, LowSNR       & 17,18     & 1.49446    & 2.11         & 0.09       & 0.38       & 1.43           & 10.3    & 7.6             \\
1400212743 & 2113.01 & pFP         & pCO, pVshape, pSS, FSCP              & 25        & 5.24612    & 4.19         & 0.10        & 0.50        & 1.76           & 10.5    & 7.5             \\
2041563029 & 1427.01 & pFP         & LCMOD, LowSNR, Long-P, FSCP          & 16        & 12.80758   & 3.60          & 2.71       & 1.02       & 0.63           & 10.1    & 3.9             \\
238061845  & 579.01  & pFP         & Vshape, pOED, pSS, UC                & 6,7,8     & 1.68418    & 1.52         & 0.07       & 1.26       & 4.94           & 9.7     & 8.0             \\
258871793  & 1843.01 & pFP         & LCMOD, FSCP, pSS, pCO, pEV           & 20        & 9.14692    & 4.04         & 0.80        & 1.02       & 1.28           & 12.6    & 5.2             \\
260271203  & 207.01  & pFP         & pTD, SS, pVshape                     & 4         & 5.649621   & 4.15         & 4.74       & 2.47       & 1.21           & 13.5    & 3.3             \\ \hline
\end{tabular}
\begin{tablenotes}
\item[†] Known planet. 
\end{tablenotes}
\end{threeparttable}
\caption{An extract of the catalog generated in this work. The TOIs shown here are drawn from the catalog such that the table contains an equal number of each disposition class. The full table is available in the Supplementary material section.}
\label{tab:catalog}
\end{table}
\end{landscape}
\section*{Acknowledgements}
This research has made use of the NASA Exoplanet Archive, which is operated by the California Institute of Technology, under contract with the National Aeronautics and Space Administration under the Exoplanet Exploration Program.
We acknowledge the use of public TESS Alert data from the pipelines at the TESS Science Office and at the TESS Science Processing Operations Center (SPOC) and from the Massachusetts Institute of Technology Quick-Look Pipeline (QLP).
This research has made use of the Exoplanet Follow-up Observation Program website, which is operated by the California Institute of Technology, under contract with the National Aeronautics and Space Administration under the Exoplanet Exploration Program.
This publication uses data generated via the Zooniverse.org platform, development of which is funded by generous support, including
a Global Impact Award from Google, and by a grant from the Alfred P. Sloan Foundation.

\textit{Software}: \texttt{DAVE} \citep{2019AJ....157..124K}, \texttt{eleanor} \citep{2019PASP..131i4502F}.

\section*{Data Availability}

The data underlying this article will be shared on reasonable request to the corresponding author.



\bibliographystyle{mnras}
\bibliography{biblio} 

\begin{thebibliography}{}
\makeatletter
\relax
\def\mn@urlcharsother{\let\do\@makeother \do\$\do\&\do\#\do\^\do\_\do\%\do\~}
\def\mn@doi{\begingroup\mn@urlcharsother \@ifnextchar [ {\mn@doi@}
  {\mn@doi@[]}}
\def\mn@doi@[#1]#2{\def\@tempa{#1}\ifx\@tempa\@empty \href
  {http://dx.doi.org/#2} {doi:#2}\else \href {http://dx.doi.org/#2} {#1}\fi
  \endgroup}
\def\mn@eprint#1#2{\mn@eprint@#1:#2::\@nil}
\def\mn@eprint@arXiv#1{\href {http://arxiv.org/abs/#1} {{\tt arXiv:#1}}}
\def\mn@eprint@dblp#1{\href {http://dblp.uni-trier.de/rec/bibtex/#1.xml}
  {dblp:#1}}
\def\mn@eprint@#1:#2:#3:#4\@nil{\def\@tempa {#1}\def\@tempb {#2}\def\@tempc
  {#3}\ifx \@tempc \@empty \let \@tempc \@tempb \let \@tempb \@tempa \fi \ifx
  \@tempb \@empty \def\@tempb {arXiv}\fi \@ifundefined
  {mn@eprint@\@tempb}{\@tempb:\@tempc}{\expandafter \expandafter \csname
  mn@eprint@\@tempb\endcsname \expandafter{\@tempc}}}

\bibitem[\protect\citeauthoryear{{Bach-M{\o}ller} \&
  {J{\o}rgensen}}{{Bach-M{\o}ller} \&
  {J{\o}rgensen}}{2021}]{2021MNRAS.500.1313B}
{Bach-M{\o}ller} N.,  {J{\o}rgensen} U.~G.,  2021, \mn@doi [\mnras]
  {10.1093/mnras/staa3321}, \href
  {https://ui.adsabs.harvard.edu/abs/2021MNRAS.500.1313B} {500, 1313}

\bibitem[\protect\citeauthoryear{{Baranne} et~al.,}{{Baranne}
  et~al.}{1996}]{1996A&AS..119..373B}
{Baranne} A.,  et~al., 1996, \aaps, \href
  {https://ui.adsabs.harvard.edu/abs/1996A&AS..119..373B} {119, 373}

\bibitem[\protect\citeauthoryear{{Barclay}, {Pepper}  \& {Quintana}}{{Barclay}
  et~al.}{2018}]{2018ApJS..239....2B}
{Barclay} T.,  {Pepper} J.,   {Quintana} E.~V.,  2018, \mn@doi [\apjs]
  {10.3847/1538-4365/aae3e9}, \href
  {https://ui.adsabs.harvard.edu/abs/2018ApJS..239....2B} {239, 2}

\bibitem[\protect\citeauthoryear{{Beaug{\'e}} \& {Nesvorn{\'y}}}{{Beaug{\'e}}
  \& {Nesvorn{\'y}}}{2013}]{2013ApJ...763...12B}
{Beaug{\'e}} C.,  {Nesvorn{\'y}} D.,  2013, \mn@doi [\apj]
  {10.1088/0004-637X/763/1/12}, \href
  {https://ui.adsabs.harvard.edu/abs/2013ApJ...763...12B} {763, 12}

\bibitem[\protect\citeauthoryear{{Borucki} et~al.,}{{Borucki}
  et~al.}{2010}]{2010Sci...327..977B}
{Borucki} W.~J.,  et~al., 2010, \mn@doi [Science] {10.1126/science.1185402},
  \href {https://ui.adsabs.harvard.edu/abs/2010Sci...327..977B} {327, 977}

\bibitem[\protect\citeauthoryear{{Cacciapuoti} et~al.,}{{Cacciapuoti}
  et~al.}{2022}]{2022MNRAS.513..102C}
{Cacciapuoti} L.,  et~al., 2022, \mn@doi [\mnras] {10.1093/mnras/stac652},
  \href {https://ui.adsabs.harvard.edu/abs/2022MNRAS.513..102C} {513, 102}

\bibitem[\protect\citeauthoryear{Christiansen et~al.,}{Christiansen
  et~al.}{2018}]{Christiansen_2018}
Christiansen J.~L.,  et~al., 2018, \mn@doi [The Astronomical Journal]
  {10.3847/1538-3881/aa9be0}, 155, 57

\bibitem[\protect\citeauthoryear{{Ciardi}, {Beichman}, {Horch}  \&
  {Howell}}{{Ciardi} et~al.}{2015}]{2015ApJ...805...16C}
{Ciardi} D.~R.,  {Beichman} C.~A.,  {Horch} E.~P.,   {Howell} S.~B.,  2015,
  \mn@doi [\apj] {10.1088/0004-637X/805/1/16}, \href
  {https://ui.adsabs.harvard.edu/abs/2015ApJ...805...16C} {805, 16}

\bibitem[\protect\citeauthoryear{{Ciardi}, {Pepper}, {Colon}, {Kane}  \&
  {Astrophysical Community}}{{Ciardi} et~al.}{2018}]{2018arXiv181008689C}
{Ciardi} D.~R.,  {Pepper} J.,  {Colon} K.,  {Kane} S.~R.,   {Astrophysical
  Community} W. I. f.~t.,  2018, arXiv e-prints, \href
  {https://ui.adsabs.harvard.edu/abs/2018arXiv181008689C} {p. arXiv:1810.08689}

\bibitem[\protect\citeauthoryear{{Coughlin} et~al.,}{{Coughlin}
  et~al.}{2016}]{2016ApJS..224...12C}
{Coughlin} J.~L.,  et~al., 2016, \mn@doi [\apjs] {10.3847/0067-0049/224/1/12},
  \href {https://ui.adsabs.harvard.edu/abs/2016ApJS..224...12C} {224, 12}

\bibitem[\protect\citeauthoryear{Cox, Oh, Simmons, Lintott, Masters, Greenhill,
  Graham  \& Holmes}{Cox et~al.}{2015}]{7106413}
Cox J.,  Oh E.~Y.,  Simmons B.,  Lintott C.,  Masters K.,  Greenhill A.,
  Graham G.,   Holmes K.,  2015, \mn@doi [Computing in Science & Engineering]
  {10.1109/MCSE.2015.65}, 17, 28

\bibitem[\protect\citeauthoryear{{Eisner}}{{Eisner}}{2022}]{2022ascl.soft05006E}
{Eisner} N.~L.,  2022, {LATTE: Lightcurve Analysis Tool for Transiting
  Exoplanet}, Astrophysics Source Code Library, record ascl:2205.006
  (\mn@eprint {ascl} {2205.006})

\bibitem[\protect\citeauthoryear{{Eisner} et~al.,}{{Eisner}
  et~al.}{2020}]{2020MNRAS.494..750E}
{Eisner} N.~L.,  et~al., 2020, \mn@doi [\mnras] {10.1093/mnras/staa138}, \href
  {https://ui.adsabs.harvard.edu/abs/2020MNRAS.494..750E} {494, 750}

\bibitem[\protect\citeauthoryear{{Eisner} et~al.,}{{Eisner}
  et~al.}{2021}]{2021MNRAS.501.4669E}
{Eisner} N.~L.,  et~al., 2021, \mn@doi [\mnras] {10.1093/mnras/staa3739}, \href
  {https://ui.adsabs.harvard.edu/abs/2021MNRAS.501.4669E} {501, 4669}

\bibitem[\protect\citeauthoryear{{Eisner}, {Lintott}, {Aigrain}, {Barragan}  \&
  {Nicholson}}{{Eisner} et~al.}{2022a}]{2022BAAS...54e.414E}
{Eisner} N.,  {Lintott} C.,  {Aigrain} S.,  {Barragan} O.,   {Nicholson} B.,
  2022a, in Bulletin of the American Astronomical Society. p. 102.414

\bibitem[\protect\citeauthoryear{{Eisner} et~al.,}{{Eisner}
  et~al.}{2022b}]{2022MNRAS.511.4710E}
{Eisner} N.~L.,  et~al., 2022b, \mn@doi [\mnras] {10.1093/mnras/stab3619},
  \href {https://ui.adsabs.harvard.edu/abs/2022MNRAS.511.4710E} {511, 4710}

\bibitem[\protect\citeauthoryear{{Faigler} \& {Mazeh}}{{Faigler} \&
  {Mazeh}}{2011}]{2011MNRAS.415.3921F}
{Faigler} S.,  {Mazeh} T.,  2011, \mn@doi [\mnras]
  {10.1111/j.1365-2966.2011.19011.x}, \href
  {https://ui.adsabs.harvard.edu/abs/2011MNRAS.415.3921F} {415, 3921}

\bibitem[\protect\citeauthoryear{{Feinstein} et~al.,}{{Feinstein}
  et~al.}{2019}]{2019PASP..131i4502F}
{Feinstein} A.~D.,  et~al., 2019, \mn@doi [\pasp] {10.1088/1538-3873/ab291c},
  \href {https://ui.adsabs.harvard.edu/abs/2019PASP..131i4502F} {131, 094502}

\bibitem[\protect\citeauthoryear{{Fiscale} et~al.,}{{Fiscale}
  et~al.}{2021}]{Fiscale21}
{Fiscale} S.,  et~al., 2021, \mn@doi [Research Notes of the American
  Astronomical Society] {10.3847/2515-5172/abf56b}, \href
  {https://ui.adsabs.harvard.edu/abs/2021RNAAS...5...91F} {5, 91}

\bibitem[\protect\citeauthoryear{{Fiscale} et~al.,}{{Fiscale}
  et~al.}{2023}]{Fiscale22}
{Fiscale} S.,  et~al., 2023, Lecture Notes in Computer Science (LNCS),
  Springer, accepted for publication

\bibitem[\protect\citeauthoryear{{Fischer} et~al.,}{{Fischer}
  et~al.}{2012}]{2012MNRAS.419.2900F}
{Fischer} D.~A.,  et~al., 2012, \mn@doi [\mnras]
  {10.1111/j.1365-2966.2011.19932.x}, \href
  {https://ui.adsabs.harvard.edu/abs/2012MNRAS.419.2900F} {419, 2900}

\bibitem[\protect\citeauthoryear{{Gaia Collaboration} et~al.,}{{Gaia
  Collaboration} et~al.}{2021}]{2021A&A...649A...1G}
{Gaia Collaboration} et~al., 2021, \mn@doi [\aap]
  {10.1051/0004-6361/202039657}, \href
  {https://ui.adsabs.harvard.edu/abs/2021A&A...649A...1G} {649, A1}

\bibitem[\protect\citeauthoryear{{Gangestad}, {Henning}, {Persinger}  \&
  {Ricker}}{{Gangestad} et~al.}{2013}]{2013arXiv1306.5333G}
{Gangestad} J.~W.,  {Henning} G.~A.,  {Persinger} R.~R.,   {Ricker} G.~R.,
  2013, arXiv e-prints, \href
  {https://ui.adsabs.harvard.edu/abs/2013arXiv1306.5333G} {p. arXiv:1306.5333}

\bibitem[\protect\citeauthoryear{{Giacalone} et~al.,}{{Giacalone}
  et~al.}{2021}]{2021AJ....161...24G}
{Giacalone} S.,  et~al., 2021, \mn@doi [\aj] {10.3847/1538-3881/abc6af}, \href
  {https://ui.adsabs.harvard.edu/abs/2021AJ....161...24G} {161, 24}

\bibitem[\protect\citeauthoryear{{Gilbert} et~al.,}{{Gilbert}
  et~al.}{2020}]{2020AJ....160..116G}
{Gilbert} E.~A.,  et~al., 2020, \mn@doi [\aj] {10.3847/1538-3881/aba4b2}, \href
  {https://ui.adsabs.harvard.edu/abs/2020AJ....160..116G} {160, 116}

\bibitem[\protect\citeauthoryear{{Hsu}, {Ford}, {Ragozzine}  \& {Ashby}}{{Hsu}
  et~al.}{2019}]{Hsuetal}
{Hsu} D.~C.,  {Ford} E.~B.,  {Ragozzine} D.,   {Ashby} K.,  2019, \mn@doi [\aj]
  {10.3847/1538-3881/ab31ab}, \href
  {https://ui.adsabs.harvard.edu/abs/2019AJ....158..109H} {158, 109}

\bibitem[\protect\citeauthoryear{{Huang} et~al.,}{{Huang}
  et~al.}{2020}]{2020RNAAS...4..204H}
{Huang} C.~X.,  et~al., 2020, \mn@doi [Research Notes of the American
  Astronomical Society] {10.3847/2515-5172/abca2e}, \href
  {https://ui.adsabs.harvard.edu/abs/2020RNAAS...4..204H} {4, 204}

\bibitem[\protect\citeauthoryear{{Jenkins} et~al.,}{{Jenkins}
  et~al.}{2016}]{2016SPIE.9913E..3EJ}
{Jenkins} J.~M.,  et~al., 2016, in {Chiozzi} G.,  {Guzman} J.~C.,  eds,
  Society of Photo-Optical Instrumentation Engineers (SPIE) Conference Series
  Vol. 9913, Software and Cyberinfrastructure for Astronomy IV. p. 99133E,
  \mn@doi{10.1117/12.2233418}

\bibitem[\protect\citeauthoryear{{Kopparapu} et~al.,}{{Kopparapu}
  et~al.}{2013}]{2013ApJ...765..131K}
{Kopparapu} R.~K.,  et~al., 2013, \mn@doi [\apj] {10.1088/0004-637X/765/2/131},
  \href {https://ui.adsabs.harvard.edu/abs/2013ApJ...765..131K} {765, 131}

\bibitem[\protect\citeauthoryear{{Kostov} et~al.,}{{Kostov}
  et~al.}{2019a}]{2019AJ....157..124K}
{Kostov} V.~B.,  et~al., 2019a, \mn@doi [\aj] {10.3847/1538-3881/ab0110}, \href
  {https://ui.adsabs.harvard.edu/abs/2019AJ....157..124K} {157, 124}

\bibitem[\protect\citeauthoryear{{Kostov} et~al.,}{{Kostov}
  et~al.}{2019b}]{2019AJ....158...32K}
{Kostov} V.~B.,  et~al., 2019b, \mn@doi [\aj] {10.3847/1538-3881/ab2459}, \href
  {https://ui.adsabs.harvard.edu/abs/2019AJ....158...32K} {158, 32}

\bibitem[\protect\citeauthoryear{{Kostov} et~al.,}{{Kostov}
  et~al.}{2022}]{2022PASP..134d4401K}
{Kostov} V.~B.,  et~al., 2022, \mn@doi [\pasp] {10.1088/1538-3873/ac5de0},
  \href {https://ui.adsabs.harvard.edu/abs/2022PASP..134d4401K} {134, 044401}

\bibitem[\protect\citeauthoryear{{Kristiansen} et~al.,}{{Kristiansen}
  et~al.}{2022}]{2022PASP..134g4401K}
{Kristiansen} M. H.~K.,  et~al., 2022, \mn@doi [\pasp]
  {10.1088/1538-3873/ac6e06}, \href
  {https://ui.adsabs.harvard.edu/abs/2022PASP..134g4401K} {134, 074401}

\bibitem[\protect\citeauthoryear{Kuchner et~al.,}{Kuchner
  et~al.}{2016}]{Kuchner_2016}
Kuchner M.~J.,  et~al., 2016, \mn@doi [The Astrophysical Journal]
  {10.3847/0004-637X/830/2/84}, 830, 84

\bibitem[\protect\citeauthoryear{{Lintott} et~al.,}{{Lintott}
  et~al.}{2008}]{2008MNRAS.389.1179L}
{Lintott} C.~J.,  et~al., 2008, \mn@doi [\mnras]
  {10.1111/j.1365-2966.2008.13689.x}, \href
  {https://ui.adsabs.harvard.edu/abs/2008MNRAS.389.1179L} {389, 1179}

\bibitem[\protect\citeauthoryear{{Lissauer}, {Rowe}, {Jontof-Hutter}  \&
  {Fabrycky}}{{Lissauer} et~al.}{2022}]{2022LPICo2687.3027L}
{Lissauer} J.~J.,  {Rowe} J.~F.,  {Jontof-Hutter} D.,   {Fabrycky} D.~C.,
  2022, in LPI Contributions. p.~3027

\bibitem[\protect\citeauthoryear{{Lomb}}{{Lomb}}{1976}]{1976Ap&SS..39..447L}
{Lomb} N.~R.,  1976, \mn@doi [\apss] {10.1007/BF00648343}, \href
  {https://ui.adsabs.harvard.edu/abs/1976Ap&SS..39..447L} {39, 447}

\bibitem[\protect\citeauthoryear{{Magliano} et~al.,}{{Magliano}
  et~al.}{2022}]{Magliano2022}
{Magliano} C.,  et~al., 2022, \mn@doi [\mnras] {10.1093/mnras/stac3404}, \href
  {https://ui.adsabs.harvard.edu/abs/2022MNRAS.tmp.3168M} {}

\bibitem[\protect\citeauthoryear{{Mann} et~al.,}{{Mann}
  et~al.}{2022}]{2022AJ....163..156M}
{Mann} A.~W.,  et~al., 2022, \mn@doi [\aj] {10.3847/1538-3881/ac511d}, \href
  {https://ui.adsabs.harvard.edu/abs/2022AJ....163..156M} {163, 156}

\bibitem[\protect\citeauthoryear{{McCauliff} et~al.,}{{McCauliff}
  et~al.}{2015}]{2015ApJ...806....6M}
{McCauliff} S.~D.,  et~al., 2015, \mn@doi [\apj] {10.1088/0004-637X/806/1/6},
  \href {https://ui.adsabs.harvard.edu/abs/2015ApJ...806....6M} {806, 6}

\bibitem[\protect\citeauthoryear{{Mislis}, {Bachelet}, {Alsubai}, {Bramich}  \&
  {Parley}}{{Mislis} et~al.}{2016}]{2016MNRAS.455..626M}
{Mislis} D.,  {Bachelet} E.,  {Alsubai} K.~A.,  {Bramich} D.~M.,   {Parley} N.,
   2016, \mn@doi [\mnras] {10.1093/mnras/stv2333}, \href
  {https://ui.adsabs.harvard.edu/abs/2016MNRAS.455..626M} {455, 626}

\bibitem[\protect\citeauthoryear{{Morris} \& {Naftilan}}{{Morris} \&
  {Naftilan}}{1993}]{1993ApJ...419..344M}
{Morris} S.~L.,  {Naftilan} S.~A.,  1993, \mn@doi [\apj] {10.1086/173488},
  \href {https://ui.adsabs.harvard.edu/abs/1993ApJ...419..344M} {419, 344}

\bibitem[\protect\citeauthoryear{{Morton}}{{Morton}}{2012}]{2012ApJ...761....6M}
{Morton} T.~D.,  2012, \mn@doi [\apj] {10.1088/0004-637X/761/1/6}, \href
  {https://ui.adsabs.harvard.edu/abs/2012ApJ...761....6M} {761, 6}

\bibitem[\protect\citeauthoryear{{Olmschenk} et~al.,}{{Olmschenk}
  et~al.}{2021}]{Olm2021}
{Olmschenk} G.,  et~al., 2021, \mn@doi [\aj] {10.3847/1538-3881/abf4c6}, \href
  {https://ui.adsabs.harvard.edu/abs/2021AJ....161..273O} {161, 273}

\bibitem[\protect\citeauthoryear{{Pepe} et~al.,}{{Pepe}
  et~al.}{2004}]{2004A&A...423..385P}
{Pepe} F.,  et~al., 2004, \mn@doi [\aap] {10.1051/0004-6361:20040389}, \href
  {https://ui.adsabs.harvard.edu/abs/2004A&A...423..385P} {423, 385}

\bibitem[\protect\citeauthoryear{{Ricker} et~al.,}{{Ricker}
  et~al.}{2015}]{2015JATIS...1a4003R}
{Ricker} G.~R.,  et~al., 2015, \mn@doi [Journal of Astronomical Telescopes,
  Instruments, and Systems] {10.1117/1.JATIS.1.1.014003}, \href
  {https://ui.adsabs.harvard.edu/abs/2015JATIS...1a4003R} {1, 014003}

\bibitem[\protect\citeauthoryear{{Samek}, {Wiegand}  \& {M{\"u}ller}}{{Samek}
  et~al.}{2017}]{2017arXiv170808296S}
{Samek} W.,  {Wiegand} T.,   {M{\"u}ller} K.-R.,  2017, arXiv e-prints, \href
  {https://ui.adsabs.harvard.edu/abs/2017arXiv170808296S} {p. arXiv:1708.08296}

\bibitem[\protect\citeauthoryear{{Scargle}}{{Scargle}}{1982}]{1982ApJ...263..835S}
{Scargle} J.~D.,  1982, \mn@doi [\apj] {10.1086/160554}, \href
  {https://ui.adsabs.harvard.edu/abs/1982ApJ...263..835S} {263, 835}

\bibitem[\protect\citeauthoryear{{Shallue} \& {Vanderburg}}{{Shallue} \&
  {Vanderburg}}{2018}]{2018AJ....155...94S}
{Shallue} C.~J.,  {Vanderburg} A.,  2018, \mn@doi [\aj]
  {10.3847/1538-3881/aa9e09}, \href
  {https://ui.adsabs.harvard.edu/abs/2018AJ....155...94S} {155, 94}

\bibitem[\protect\citeauthoryear{{Shporer}}{{Shporer}}{2017}]{2017PASP..129g2001S}
{Shporer} A.,  2017, \mn@doi [\pasp] {10.1088/1538-3873/aa7112}, \href
  {https://ui.adsabs.harvard.edu/abs/2017PASP..129g2001S} {129, 072001}

\bibitem[\protect\citeauthoryear{{Smalley} et~al.,}{{Smalley}
  et~al.}{2011}]{2011A&A...526A.130S}
{Smalley} B.,  et~al., 2011, \mn@doi [\aap] {10.1051/0004-6361/201015992},
  \href {https://ui.adsabs.harvard.edu/abs/2011A&A...526A.130S} {526, A130}

\bibitem[\protect\citeauthoryear{{Sullivan} et~al.,}{{Sullivan}
  et~al.}{2015}]{2015ApJ...809...77S}
{Sullivan} P.~W.,  et~al., 2015, \mn@doi [\apj] {10.1088/0004-637X/809/1/77},
  \href {https://ui.adsabs.harvard.edu/abs/2015ApJ...809...77S} {809, 77}

\bibitem[\protect\citeauthoryear{{Szab{\'o}} \& {Kiss}}{{Szab{\'o}} \&
  {Kiss}}{2011}]{2011ApJ...727L..44S}
{Szab{\'o}} G.~M.,  {Kiss} L.~L.,  2011, \mn@doi [\apjl]
  {10.1088/2041-8205/727/2/L44}, \href
  {https://ui.adsabs.harvard.edu/abs/2011ApJ...727L..44S} {727, L44}

\bibitem[\protect\citeauthoryear{{Tey} et~al.,}{{Tey} et~al.}{2023}]{Tey2023}
{Tey} E.,  et~al., 2023, \mn@doi [\aj] {10.3847/1538-3881/acad85}, \href
  {https://ui.adsabs.harvard.edu/abs/2023AJ....165...95T} {165, 95}

\bibitem[\protect\citeauthoryear{{Thompson} et~al.,}{{Thompson}
  et~al.}{2018}]{2018ApJS..235...38T}
{Thompson} S.~E.,  et~al., 2018, \mn@doi [\apjs] {10.3847/1538-4365/aab4f9},
  \href {https://ui.adsabs.harvard.edu/abs/2018ApJS..235...38T} {235, 38}

\bibitem[\protect\citeauthoryear{{Valizadegan} et~al.,}{{Valizadegan}
  et~al.}{2022}]{Valizadeganetal}
{Valizadegan} H.,  et~al., 2022, \mn@doi [\apj] {10.3847/1538-4357/ac4399},
  \href {https://ui.adsabs.harvard.edu/abs/2022ApJ...926..120V} {926, 120}

\bibitem[\protect\citeauthoryear{{Visser}, {Bosma}  \& {Postma}}{{Visser}
  et~al.}{2022}]{Visser2022}
{Visser} K.,  {Bosma} B.,   {Postma} E.,  2022, \mn@doi [Journal of
  Astronomical Instrumentation] {10.1142/S2251171722500118}, \href
  {https://ui.adsabs.harvard.edu/abs/2022JAI....1150011V} {11, 2250011}

\bibitem[\protect\citeauthoryear{{Wang} et~al.,}{{Wang}
  et~al.}{2015}]{2015ApJ...815..127W}
{Wang} J.,  et~al., 2015, \mn@doi [\apj] {10.1088/0004-637X/815/2/127}, \href
  {https://ui.adsabs.harvard.edu/abs/2015ApJ...815..127W} {815, 127}

\bibitem[\protect\citeauthoryear{{Wenger} et~al.,}{{Wenger}
  et~al.}{2000}]{2000A&AS..143....9W}
{Wenger} M.,  et~al., 2000, \mn@doi [\aaps] {10.1051/aas:2000332}, \href
  {https://ui.adsabs.harvard.edu/abs/2000A&AS..143....9W} {143, 9}

\bibitem[\protect\citeauthoryear{{Yu} et~al.,}{{Yu} et~al.}{2019}]{Yu2019}
{Yu} L.,  et~al., 2019, \mn@doi [\aj] {10.3847/1538-3881/ab21d6}, \href
  {https://ui.adsabs.harvard.edu/abs/2019AJ....158...25Y} {158, 25}

\makeatother
\end{thebibliography}



\bsp	
\label{lastpage}
\end{document}